\documentclass[aip, jcp, notitlepage, floatfix, preprint,
  citeautoscript, twocolumn, nofootinbib]{revtex4-1}

\usepackage{amsmath}
\usepackage{amsfonts}
\usepackage{graphicx}
\usepackage{dcolumn}
\usepackage{miller}
\usepackage[table]{xcolor}
\usepackage[version=3]{mhchem}
\usepackage{transparent}
\usepackage{gensymb}
\usepackage{tabularx}
\usepackage{etoolbox}
\usepackage[hidelinks]{hyperref}
\usepackage{bm}
\usepackage{relsize}
\usepackage[normalem]{ulem}
\usepackage{cases}
\usepackage{siunitx, multirow}
\usepackage{tabularray}
\usepackage[left=1.5cm, right=1.5cm, top=1.785cm, bottom=2.0cm]{geometry}
\usepackage{float}

\def\maxfloatwidth{%
  \ifdim\columnwidth>246.0pt
  300.0pt  \else
  \columnwidth
  \fi
}

\newcommand{\mbf}[1]{\mathbf{#1}}

\newcommand{\tcb}[1]{\textcolor{black}{#1}}
\newcommand{\etal}{\emph{et al.}}

\definecolor{bgpeach}{rgb}{1.000,0.925,0.850}
\definecolor{fggray}{rgb}{0.384,0.435,0.471}

\begin{document}

\title{\large The limit of macroscopic 
homogeneous ice nucleation at the nanoscale}

\author{John A. Hayton} 
\affiliation{Yusuf Hamied Department of Chemistry, University of
  Cambridge, Lensfield Road, Cambridge CB2 1EW, United Kingdom}

\author{Michael B. Davies}
\affiliation{Department of Physics and Astronomy, University College
  London, London WC1E 6BT, United Kingdom}
\affiliation{Yusuf Hamied Department of Chemistry, University of
  Cambridge, Lensfield Road, Cambridge CB2 1EW, United Kingdom}

\author{Thomas F. Whale}
\affiliation{Department of Chemistry, University of Warwick, 
Gibbet Hill Road, Coventry CV4 7AL, United Kingdom.}

\author{Angelos Michaelides}
\affiliation{Yusuf Hamied Department of Chemistry, University of
  Cambridge, Lensfield Road, Cambridge CB2 1EW, United Kingdom}

\author{Stephen J. Cox}
\affiliation{Yusuf Hamied Department of Chemistry, University of
  Cambridge, Lensfield Road, Cambridge CB2 1EW, United Kingdom}
\email{sjc236@cam.ac.uk}

\date{22 June 2023}

\begin{abstract}
Nucleation in
  small volumes of water has garnered renewed interest due to the
  relevance of pore condensation and freezing under conditions of low
  partial pressures of water, such as in the upper
  troposphere. Molecular simulations can in principle provide insight
  on this process at the molecular scale that is challenging to
  achieve experimentally. However, there are discrepancies in the
  literature as to whether the rate in confined systems is enhanced or
  suppressed relative to bulk water at the same temperature and
  pressure. In this study, we investigate the extent to which the size
  of the critical nucleus and the rate at which it grows in thin films
  of water are affected by the thickness of the film. Our results
  suggest that nucleation remains bulk-like in films that are barely
  large enough accommodate a critical nucleus. This conclusion seems
  robust to the presence of physical confining boundaries. We also
  discuss the difficulties in unambiguously determining homogeneous
  nucleation rates in nanoscale systems, owing to the challenges in
  defining the volume. Our results suggest any impact on a film's
  thickness on the rate is largely inconsequential for present day
  experiments.
\end{abstract}

\maketitle

\section{Introduction}
\label{sec:Intro}

The formation of ice from liquid water is one of the most important
phase transitions on Earth, and plays a vital role in climate
science,\cite{Cantrell2005,Satoh2018,DeMott2010,Hoose2010,Gettelman2010,Gettelman2012,Hudait2016}
cryopreservation,\cite{Kiani2011, Rall1985, 2014AaCt}
geology\cite{gerrard2012rocks,ashton1986river} and many industrial
applications.\cite{Rykaczewski2013,Fukasawa2001,Fukasawa2002,Ohji2013}
For example, many properties of clouds are affected by the relative
compositions of ice and
water,\cite{baker1997cloud,carslaw2002cosmic,DeMott2010,boucher2013clouds}
and consequently, the accuracy of climate models rely heavily on
parameterizations to predict ice nucleation in the atmosphere.

Broadly speaking, when ice forms, it can do so either heterogeneously,
where the surface of, say, a solid particle facilitates nucleation, or
homogeneously, in the absence of such surfaces. Despite heterogeneous
nucleation being by far more common, homogeneous nucleation is
important when temperatures approach approx. $-40$\degree C, e.g., in
cirrus cloud formation in the upper troposphere.\cite{Karcher2002,
  Jensen1998, DeMott1998, Rogers1998, DeMott2003, Satoh2018,
  DeMott2010} Yet, even in the absence of surfaces presented by solid
particles, the finite volume occupied by the liquid means an interface
(e.g., with vapor or oil) nonetheless remains. A long-standing issue
for homogeneous nucleation has therefore been to establish whether
nucleation is enhanced close to the liquid-vapor interface, or
suppressed. Owing to the small length and fast times scales involved,
however, it is experimentally challenging to establish whether
homogeneous nucleation occurs near the interface, or in the bulk of
the fluid. 
Molecular simulations have therefore been employed by
several groups to investigate where in the liquid homogeneous
nucleation
occurs.\cite{Vrbka2006,Vrbka2007,Li2013,johnston2012crystallization,
HajiAkbari2014,HajiAkbari2017}
While pioneering simulation studies from
Jungwirth\cite{Vrbka2006,Vrbka2007} and co-workers suggested that
nucleation is enhanced at the liquid-vapor interface, their results
are likely affected by the finite size of the simulation
cell.\cite{Cox2013} The broad consensus from multiple simulation
studies is that ice formation occurs away from the interface, in
regions of the fluid that are bulk-like. \tcb{These conclusions are also 
supported by thermodynamic arguments made by Qiu and Molinero.\cite{qiu2018so} }

%\tcb{If ice forms preferentially in the bulk of the liquid, what 
%does this mean for the observed rate of ice nucleation? For the 
%case of nanodroplets, the ice nucleation characteristics of 
%volumetric rate in Ref.~\citenum{Li2013} and temperature of maximum 
%freezing rate in Ref.~\citenum{johnston2012crystallization} have 
%been previously investigated. In 2013, using forward flux sampling, 
%Li \etal{}\cite{Li2013} show that decreasing the droplet radius from 
%approx. $4.9$\,nm to approx. $2.4$\,nm, that nucleation rates at 
%230\,K were decreased relative to that of bulk water by eight orders 
%of magnitude; for radii $\gtrsim 5$\,nm, nucleation rates were virtually 
%indistinguishable from that of bulk. With the exception of the smallest 
%droplet investigated, these significant decreases in nucleation rate 
%could be well explained by the associated change in Laplace pressure. 
%In 2012, Johnston and Molinero\cite{johnston2012crystallization} 
%demonstrated that temperatures of maximum nucleation rate can also 
%be broadly explained by a modified Gibbs-Thompson equation, applied 
%to nanodroplets of radii spanning $4.7$\,nm to $1.05$\,nm.}

If ice forms preferentially in the bulk of the liquid, what does this
mean for the observed rate of ice nucleation? In 2013, using forward
flux sampling, Li \etal{}\cite{Li2013} found, upon decreasing the
droplet radius from approx. $4.9$\,nm to approx. $2.4$\,nm, that
nucleation rates at 230\,K were decreased relative to that of bulk
water by eight orders of magnitude; for radii $\gtrsim 5$\,nm,
nucleation rates were virtually indistinguishable from that of
bulk. With the exception of the smallest droplet investigated, this
significant decrease of the rate upon decreasing the radius appeared
to be explained well by the associated change in Laplace
pressure. \tcb{These findings were consistent with a previous study by
  Johnston and Molinero.\cite{johnston2012crystallization} } However,
subsequent simulation studies that investigated ice nucleation in thin
water films,\cite{HajiAkbari2014,Lu2013} whose planar interfaces
correspond to zero Laplace pressure, found that nucleation rates were
noticeably suppressed relative to that of bulk, even for film
thicknesses $\gtrsim 5$\,nm. Owing to the computationally demanding
nature of simulating ice formation, these studies used a
coarse-grained representation of water's interactions, the mW
model.\cite{Molinero2009} In this model, instead of explicitly
representing the hydrogen bond network, the local tetrahedrality is
enforced by a three body contribution to the potential energy
function. Despite its simplicity, the mW model reasonably describes
the anomalies and structure of water and its phase behavior, including
the density maximum, the increase in heat capacity and compressibility
of the supercooled region and melting temperatures of both hexagonal
and cubic
ice.\cite{Molinero2009,moore2011structural,Limmer2011,moore2010freezing}

To add further complication to the above apparent discrepancy between
droplets and films, tour-de-force simulations employing the TIP4P/ice
model,\cite{Abascal2005} which unlike the coarse grained mW model,
accounts explicitly for electrostatic interactions and water's
hydrogen bond network, found that nucleation in thin water films was
\emph{increased} relative to that of bulk water by approx. seven
orders of magnitude.\cite{HajiAkbari2017} This increase
was attributed to an increase in average crystalline order and
cage-like structure for TIP4P/ice (in the liquid state) in a thin-film
geometry, which persisted to film thicknesses even as large as
9\,nm. These pronounced structural changes were not observed with the
coarse grained mW model. While the origin of such a discrepancy might
be attributed to a lack of essential physics in the coarse grained
model, this may raise a question mark over the use of mW to
investigate ice formation in confined geometries, despite its 
success,\cite{Bi2017,Rosky2023,moore2010freezing}
e.g., in explaining ice nucleation from vapor via a pore condensation
and freezing mechanism.\cite{David2019}

In this article, we first aim to resolve this apparent discrepancy
between the coarse grained mW model and the all-atom TIP4P/ice
model. Then, in Sec.~\ref{sec:Thin_Films}, using the computationally
more efficient mW model, we directly assess how the simplest confining
geometry of all---a thin film of water in contact with its
vapor---impacts ice formation.  In Sec.~\ref{sec:PhysicalBoundaries}
we extend our investigation to a more realistic case where water is
confined between two solid surfaces. We conclude our findings in
Sec.~\ref{sec:Concl}. A broad overview of the methodology is described
throughout the manuscript, with full details provided in
Sec.~\ref{sec:Methods} and the Electronic Supporting Information
(ESI).

\section{Structural properties of thin water films converge to their bulk values on a microscopic length scale}
\label{subsec:Justifying_mW}

As discussed above, previous work from Haji-Akbari and
Debenedetti,\cite{HajiAkbari2017} using the TIP4P/ice water model at
230\,K has reported that the nucleation rate $J(W)$ in a film of
thickness $W\approx 4$\,nm is roughly seven orders of magnitude larger
than in bulk water, whose rate we denote $J(\infty)$. This significant
enhancement of the rate has been attributed to a structure of liquid
water ``deep'' in the interior of thin films that is distinctly different
from that found for homogeneous bulk water. In particular, measures of
local order, as prescribed, e.g., by a Steinhardt order
parameter\cite{Steinhardt1983} for the $i^{\mathrm{th}}$ water
molecule,
\begin{equation}
  \label{Eqn:Steinhardt_barq6}
  \bar{Q}_6^{(i)} = \frac{Q_{6}^{(i)} + \sideset{}{^\prime}\sum_{j} Q_{6}^{(j)}}{\nu^{(i)}+1},  
\end{equation}
indicate that the average structure of liquid water is more ice-like
for thin films when compared to homogeneous bulk water under the same
conditions.\footnote{There a several similar measures of local order
in the literature,\cite{Li2011,Lechner2008,rein1996numerical} but the
analysis we present in this section is robust to the exact choice. }
The prime indicates that the sum only includes the $\nu^{(i)}$ nearest
neighbors of molecule $i$ (see ESI),
\begin{equation}
  Q_6^{(i)} = \frac{1}{\nu^{(i)}}\sqrt{ \sum_{m=-6}^{6}
  \sideset{}{^\prime}\sum_{j,k} Y_{6m}^{\ast} (\hat{\mbf{r}}_{ij})Y_{6m} (\hat{\mbf{r}}_{ik})},
\end{equation}
where $\hat{\mbf{r}}_{ij}$ is the unit vector pointing from molecule
$i$ to molecule $j$, and $Y_{6m}$ is the $m^{\rm th}$ component of a
sixth-rank spherical harmonic.
Similarly, the number of cage-like structures was also found to be
increased in thin water films. In contrast, when using the mW
potential, such measures of local structure converge to their bulk
values within approx. 1\,nm of the interface.

In a recent study, Atherton \etal{}\cite{Atherton2022} discussed the
sensitivity of the melting point of simple point charge models such as
TIP4P/ice to the choice of cutoff, $r_{\ast}$, for the
non-electrostatic interactions between water molecules. There it was
shown that decreasing $r_{\ast}$ led to a systematic increase in the
melting temperature. More importantly, it was noticed that choices of
$r_{\ast}$ typically used in molecular simulations effectively
correspond to negative pressures of a few hundred bar, when compared
to homogeneous bulk systems that use a mean-field treatment to correct
for truncated interactions. This observation was argued to be
particularly relevant when comparing nucleation in thin water films
(where standard mean-field treatments of truncated interactions have
no impact) to homogeneous bulk systems. In Ref.~\citenum{Atherton2022}, rough
arguments based on results from Bianco \etal{}\cite{Bianco2021} were used to suggest
that this effective negative pressure was the root of faster ice
nucleation in thin films, but firm evidence was lacking.

For a thin film of water in coexistence with its vapor, it can readily
be argued that the pressure far from the interfaces is approximately
0\,bar.\cite{rowlinson2013molecular} Therefore, to obtain a suitable
reference for the local structure in the homogeneous bulk fluid, we
have performed simulations in the isothermal-isobaric ensemble at a
temperature $T=230$\,K, and pressure $p = 0$\,bar (see ESI for
full simulation details), and computed
\begin{equation}
  \langle\bar{q}_6\rangle/\bar{\rho} = \frac{1}{N}\big\langle\sum_i\bar{Q}_6^{(i)}\big\rangle,
\end{equation}
where $\bar{\rho}$ is the density of bulk water, $N$ is the number of
water molecules in the simulation, and angled brackets indicate an
ensemble average. An important detail is that, when using the
TIP4P/ice potential, we have truncated and shifted non-electrostatic
interactions at $r_{\ast} = 8.5\,\text{\AA}$; following
Ref.~\citenum{Atherton2022} we indicate this molecular model as
TIP4P/ice$^{(8.5)}$. We find $\langle\bar{q}_6\rangle/\bar{\rho}
\approx 0.266$.

%We find $\langle\bar{q}_6\rangle = 8.238$\,nm$^{-3}$.

\begin{figure}[h]
    \centering
    \includegraphics[width=7.5cm]{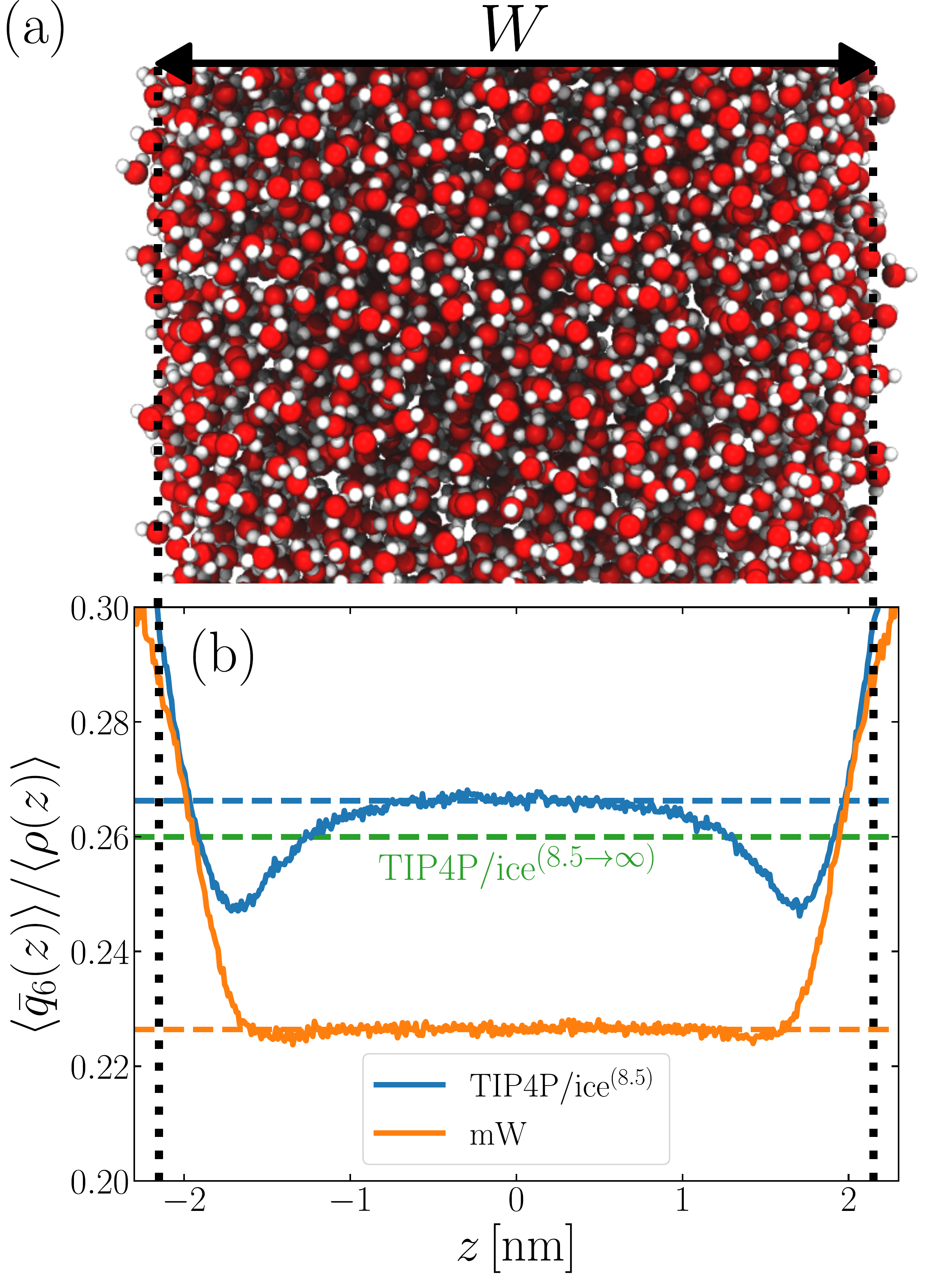}
    \caption{Structural properties of TIP4P/ice$^{(8.5)}$ and mW
      converge to their bulk values within approx. 1--1.5\,nm of the
      liquid-vapor interface. (a) Snapshot of a TIP4P/ice$^{(8.5)}$
      film of thickness $W\approx 4.3$\,nm at $T=230$\,K. The $z$ axis
      (normal to the liquid-vapor interface) lies along the
      horizontal. Local structure away from the interface soon becomes
      bulk-like, as shown by $\langle
      \bar{q}_6(z)\rangle/\langle\rho(z)\rangle$ in (b) for
      TIP4P/ice$^{(8.5)}$ (blue) and mW (orange). Dashed lines show
      $\langle \bar{q}_6\rangle/\bar{\rho}$ obtained from simulations
      of the bulk fluid at $T=230$\,K and $p=0$\,bar. When truncated
      interactions are accounted for in a mean-field fashion, a
      discrepancy in local structure in the center of the film and
      homogeneous bulk water emerges, as indicated by the dashed green
      line. The vertical dotted lines indicate the location of the
      liquid-vapor interface at $z=\pm W/2$.}
    \label{fig:Q6_Profiles}
\end{figure}

We now turn our attention to a thin film, as shown in
Fig.~\ref{fig:Q6_Profiles}a, also at $T=230$\,K. The specific system
we consider comprises 3072 TIP4P/ice$^{(8.5)}$ water molecules, and
the lateral dimensions of the simulation cell are chosen such that
$W\approx 4.3$\,nm. Throughout this paper, the width of a film is
defined as $W=N/A\bar{\rho}$,
%where $N$ is the number of molecules in the film,
where $A$ is the cross-sectional area. $W$ is then varied by changing
$A$. This simple measure is roughly consistent with the separation of
Gibbs dividing surfaces.

To analyze the spatial variation in the structure
of this film, we compute
\begin{equation}
  \langle \bar{q}_6(z)\rangle/\langle\rho(z)\rangle = \frac{1}{A\langle\rho(z)\rangle}\big\langle \sum_i
  \bar{Q}_6^{(i)}\delta(z-z_i)\big\rangle,
  \label{eqn:Q6_Profile}
\end{equation}
where $z$ is the coordinate normal to the average liquid-vapor
interface, $z_i$ is the $z$ coordinate of $i^{\rm th}$ water
molecule's oxygen atom, and
\begin{equation}
  \langle\rho(z)\rangle = \frac{1}{A}\big\langle\sum_i\delta(z-z_i)\big\rangle
\end{equation}
is the number density profile. The result of this analysis is shown in
Fig.~\ref{fig:Q6_Profiles}b;
%this profile is largely reminiscent of a
%standard number density profile, varying smoothly from
%$\langle\bar{q}_6(z)\rangle \approx 0$\,nm$^{-3}$ in the vapor phase
%to $\langle\bar{q}_6(z)\rangle \approx 8.3$\,nm$^{-3}$ in the
%liquid.
While some interesting structure is observed close to the interface on
the liquid side, the important point is that $\langle
  \bar{q}_6(z)\rangle/\langle\rho(z)\rangle \approx 0.266$ converges to
its bulk value within approx. 1--1.5\,nm of the interface, as indicated
by the dashed blue line in Fig.~\ref{fig:Q6_Profiles}b. In
  addition, the dashed green line shows
  $\langle\bar{q}_6\rangle/\langle \rho(z)\rangle \approx 0.260$
  obtained from a simulation of the bulk fluid at $T=230$\,K and
  $p=0$\,bar in which a mean-field treatment for truncated
  interactions has been employed; we denote this molecular model
  TIP4P/ice$^{(8.5\to\infty)}$. We see that the result obtained with
  TIP4P/ice$^{(8.5\to\infty)}$ lies, on average, below
  $\langle\bar{q}_6(z)\rangle/\langle\rho(z)\rangle$ obtained with
  TIP4P/ice$^{(8.5)}$.

Also shown in Fig.~\ref{fig:Q6_Profiles}b is a similar analysis for
the mW model. While we see quantitative differences with the
TIP4P/ice$^{(8.5)}$ result, the important point is that
$\langle\bar{q}_6\rangle/\langle\rho(z)\rangle \approx 0.226$
converges to its bulk value in the center of the film, on a similar
length scale to TIP4P/ice$^{(8.5)}$. This result demonstrates that the
structural differences between mW and TIP4P/ice that lead to
apparently qualitative differences in nucleation rates can be resolved
through a consistent treatment of truncated interactions.  Combined
with previous observations that the mechanism of ice formation in thin
films is similar between the two
models,\cite{HajiAkbari2014,HajiAkbari2017,Li2013} the results in this
section strongly suggest that the coarse-grained mW model likely
contains the essential physics to describe ice nucleation in these
systems.

In the following section, we will exploit the computational efficiency
of the mW model to explore how the nucleation rate in thin water films
depends upon $W$. Specifically, using the seeding technique, we will
investigate how both the size of the critical nucleus and its growth
depend upon $W$. Preliminary results for TIP4P/ice$^{(8.5)}$ with
$W=4.3$\,nm are given in the ESI.

\section{Nucleation in thin water films remains bulk-like down to very small length scales}
\label{sec:Thin_Films}

\begin{figure}[h]
    \centering \includegraphics[width=7.5cm]{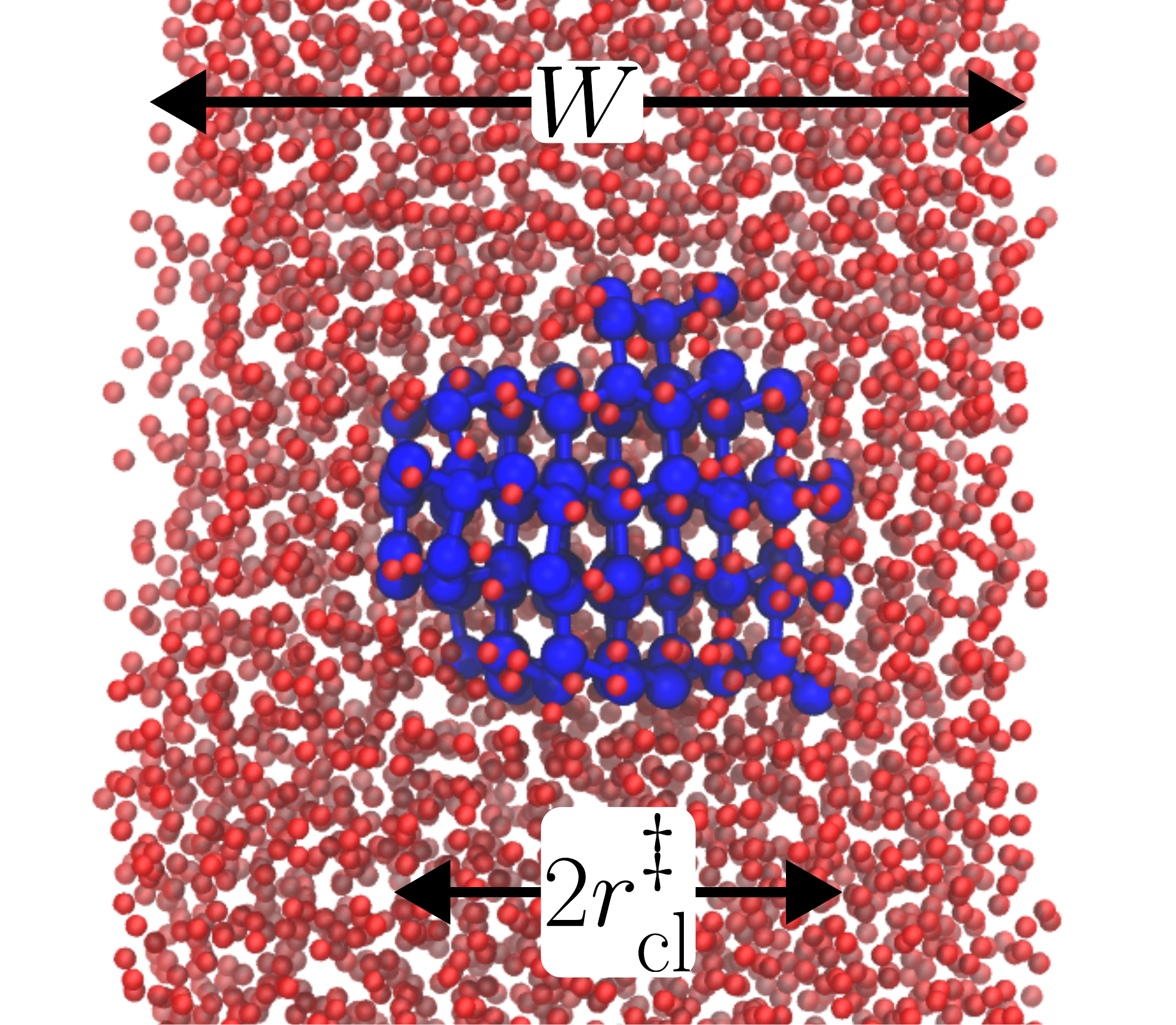}
    \caption{A snapshot from a seeding simulation with $W = 3.5$\,nm
      after initial equilibration at $T=220$\,K, using the mW
      model. Molecules identified as belonging to the largest ice-like
      cluster are shown by large blue spheres, while all other
      molecules are shown by small red spheres. In this case, the
      initial seed happens to be critical, with approximate radius
      $r_{\rm cl}^\ddagger$, as indicated.}
    \label{fig:Seeding_Snapshots}
\end{figure}

\begin{figure*}[t]
  \centering
  \includegraphics[width=16cm]{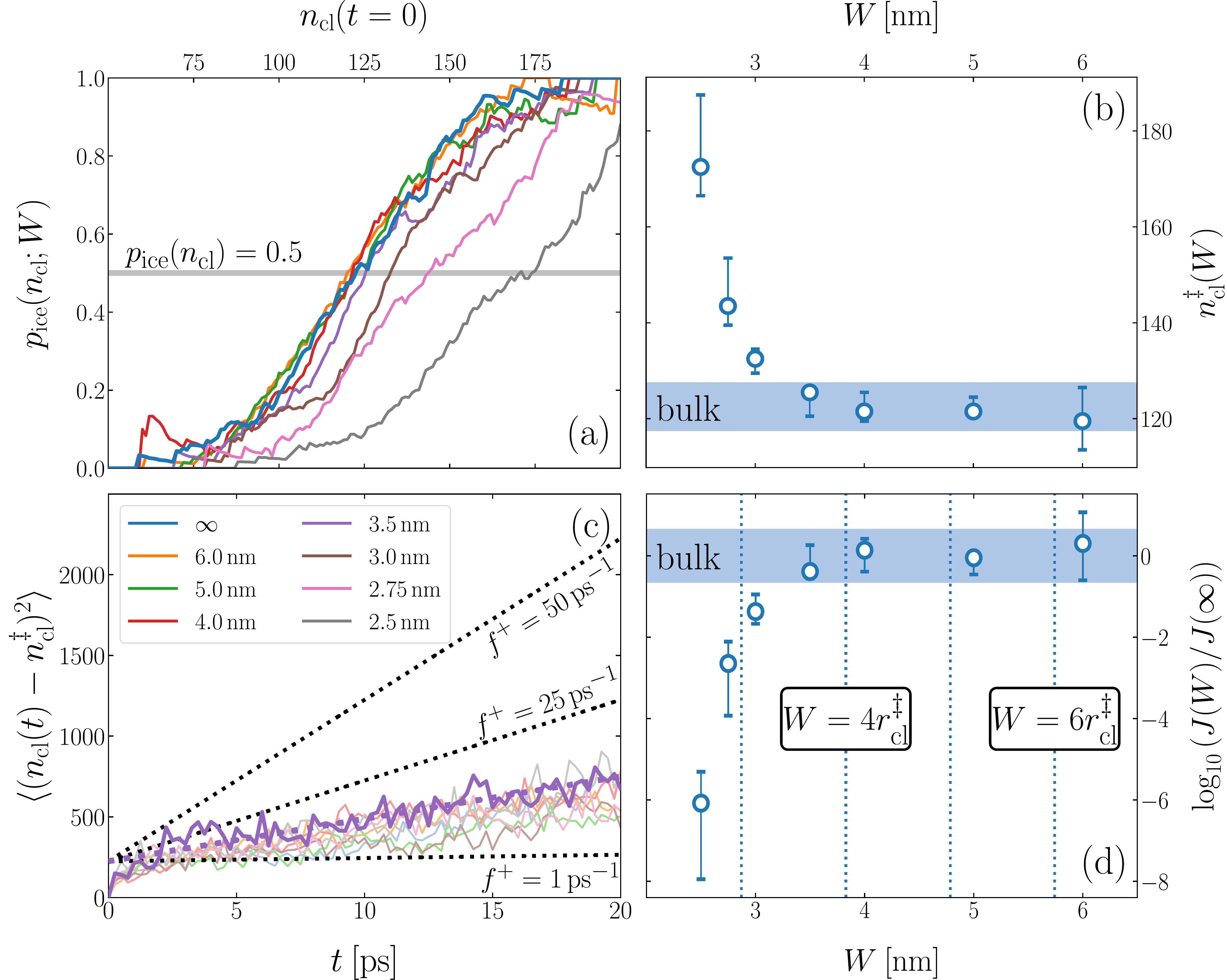}
  \caption{Assessing the impact of film width, $W$, on ice
    nucleation. (a) The size of the critical cluster for a given $W$
    [as indicated in the legend in (c)] is obtained by finding the
    size of cluster after initial equilibration, $n_{\rm cl}(t=0) =
    n_{\rm cl}^\ddagger$, for which there is equal probability to grow
    or shrink, $p_{\rm ice}(n_{\rm cl}; W) = 0.5$. We find that
    $n_{\rm cl}^\ddagger$ only differs significantly from its value in
    bulk water for $W\lesssim 3.5$\,nm, as shown in (b). Errors bars
    indicate the range of results obtained by splitting each data set
    into three. In (c) we show $\langle (n_{\rm cl}(t)-n_{\rm
      cl}^\ddagger)^2\rangle$ for each $W$. The attachment frequency
    $f^+$ is obtained by fitting a straight line after an initial
    transient period, as shown in bold for $W=3.5$\,nm. Dotted lines
    indicate hypothetical values of $f^+$ to give a sense of
    scale. Using the obtained $n_{\rm cl}^\ddagger$ and $f^+$, and
    $\Delta\mu/k_{\rm B} = 122$\,K, the nucleation rate is obtained
    from Eq.~\ref{eqn:JCNT}, as shown in (d). In panels (b) and (d),
    the shaded region indicates the result obtained from simulations
    of bulk water.}
  \label{fig:Seeding_Figures}
\end{figure*}

Having established consistency between TIP4P/ice$^{(8.5)}$ and mW in
describing the structure of thin water films, we will now investigate
the extent to which $W$ impacts ice nucleation. To do so, we will make
use of the `seeding' method, first introduced by Bai and
Li\cite{Bai2005,Bai2006} in their study of Lennard-Jones particles,
and popularized for ice nucleation by Vega and co-workers.
\cite{Sanz2013,Espinosa2016,Zaragoza2015,Espinosa2014,
  Espinosa2016B,Espinosa2016C,Espinosa2018,Bianco2021} As the seeding
approach has been detailed previously by others,
\cite{Sanz2013,Espinosa2014,Espinosa2016,Espinosa2016B} here we will
only briefly sketch an outline of the procedure.

\subsection{Investigating ice nucleation in thin water films with seeding simulations}

The principal idea behind the seeding method is to initialize the
system with a preformed ice nucleus (a `seed') and observe whether the
seed, on average, grows such that the system ends up as ice, or
shrinks such that the system ends up as liquid water. If the seed is
smaller than the critical ice nucleus, it will tend to melt, whereas
if it is larger, it will tend to grow; the size of the critical ice
nucleus can be determined from seeds that have equal tendency to grow
or melt. In our case, this was achieved by first equilibrating a bulk
crystal of hexagonal ice comprising 16000 molecules at 220\,K and
0\,bar, and then carving out a spherical cluster. The cluster was then
inserted into the center of a thin film of liquid water of thickness
$W$, which itself had been equilibrated at 220\,K; this was done after
first removing water molecules to create a spherical cavity of an
appropriate size to accommodate the ice cluster. With the molecules in
the ice cluster held fixed, the surrounding fluid was then relaxed by
performing a short 80\,ps molecular dynamics simulation. We
subsequently calculated the size of the largest cluster of ice-like
molecules in the system, $n_{\rm cl}$, which we took to define the
size of our initial ice seed. An example of such a seed is shown in
Fig.~\ref{fig:Seeding_Snapshots}.

\SetTblrInner{rowsep=4pt}

\begin{table*}
    \centering
    \begin{tabular*}{\textwidth}{@{\extracolsep{\fill}}llll}
        \hline
         $W\,\mathrm{[nm]}$  & $n_{\mathrm{cl}}^{\ddag}$ & $f^{+}\,\mathrm{[ps^{-1}]}$ & $\log_{10}\left( J/(\mathrm{m^{-3}\:s^{-1}}) \right)$ \\
         \hline
         $\infty$ & $122.5\:(117.5 - 127.5)$ & $13.7$ & $25.1\:(24.4 - 25.7)$ \\
         6.0 & $119.5\:(113.5 - 126.5)$ & $11.9$ & $25.4\:(24.5 - 26.2)$ \\
         5.0 & $121.5\:(121.5 - 124.5)$ & $9.3$ & $25.1\:(24.6 - 25.1)$ \\
         4.0 & $121.5\:(119.5 - 125.5)$ & $14.2$ & $25.2\:(24.7 - 25.5)$ \\
         3.5 & $125.5\:(120.5 - 126.5)$ & $13.1$ & $24.7\:(24.5 - 25.4)$ \\
         3.0 & $132.5\:(129.5 - 134.5)$ & $9.7$ & $23.7\:(23.4 - 24.1)$ \\
         2.75 & $143.5\:(139.5 - 153.5)$ & $9.5$ & $22.4\:(21.1 - 22.9)$ \\
         2.5 & $172.5\:(166.5 - 187.5)$ & $14.2$ & $19.0\:(17.1 - 19.8)$ \\
         \hline
    \end{tabular*}
    \caption{Summary of results from seeding simulations. For each
      $W$, the range of critical cluster sizes (indicated in
      parentheses) is obtained by splitting its data set into
      three. The range for $J(W)$ is calculated by using the largest
      and smallest values for $n_\mathrm{cl}^{\ddag}$ for a given $W$
      in Eq.~\ref{eqn:JCNT} ($\Delta\mu/k_{\rm B} = 122$\,K and
      $\bar{\rho} = 33.3774$\,nm$^{-3}$), and is also reported in
      parentheses}
    \label{tab:Results_Table}
\end{table*}

We have investigated film thicknesses in the range $2.5\,\mathrm{nm}
\lesssim W \lesssim 6\,\mathrm{nm}$, with $N\approx 6000$ throughout
(i.e., different thicknesses are achieved by varying $A$). For each
$W$, approximately 700-900 seeding simulations were performed. The
range of initial cluster sizes depended on film width, with $60
\lesssim n_{\rm cl} \lesssim 220$ overall; this was sufficient to span
both pre- and post-critical cluster sizes. In
Fig.~\ref{fig:Seeding_Figures}a, we present the probability $p_{\rm
  ice}(n_{\rm cl}; W)$ that a seed of size $n_{\rm cl}$ goes on to
form ice. The critical cluster size $n_{\rm cl}^\ddagger$ is estimated
from $p_{\rm ice}(n_{\rm cl}^\ddagger; W) = 0.5$. We observe that for
$W \gtrsim 3.5$\,nm, $n_{\rm cl}^{\ddagger} \approx 120$, which
compares well to the value obtained in a bulk simulation of mW. To
help give this result some perspective, the radius of these critical
clusters is $r_{\rm cl}^\ddagger \approx 0.96$\,nm. If we consider
$W=3.5$\,nm, this gives $(W-2r_{\rm cl}^\ddagger)/2 \approx 0.8$\,nm,
which is broadly in line with accepted thicknesses of the liquid-vapor
interface.\cite{HajiAkbari2014,Li2013,sega2016} In other words, as soon as the
films are thick enough to accommodate a bulk-like region, critical
nuclei appear ambivalent to the nearby presence of the liquid-vapor
interface. For $W\lesssim 3$\,nm, deviations of $n_{\rm cl}^\ddagger$
from its bulk value are observed, with $n_{\rm cl}^\ddagger\approx
170$ in the thinnest film investigated ($W \approx 2.5$\,nm). For all
systems investigated, $n_{\rm cl}^\ddagger$ vs. $W$ is plotted in
Fig.~\ref{fig:Seeding_Figures}b.

Computation of nucleation rates with the seeding approach relies upon
classical nucleation theory 
(CNT):\cite{Sanz2013,Espinosa2016,Bai2005,Bai2006}
\begin{equation}
  \label{eqn:JCNT}
  J = \mathcal{Z}f^{+}\bar{\rho}\exp\big(-\beta|\Delta\mu|n_{\rm cl}^\ddagger/2 \big),
\end{equation}
where $\Delta\mu$ is the chemical potential difference between bulk
ice and liquid water, $\beta = 1/k_{\rm B}T$ ($k_{\rm B}$ is
Boltzmann's constant), $\mathcal{Z} = (\beta|\Delta\mu|/6\pi n_{\rm
  cl}^\ddagger)^{1/2}$ is the Zeldovich factor, and $f^{+}$ is the
attachment frequency. For the chemical potential difference, a range
of values spanning approx. $118\,\mathrm{K} \lesssim
|\Delta\mu|/k_{\rm B} \lesssim 126\,\mathrm{K}$ have been reported in
the literature \cite{Reinhardt2012,Jacobson2009} and for simplicity,
we take $|\Delta\mu|/k_{\rm B} = 122$\,K. (We have repeated the
following analyses with the extremal values of this range, and find that the results are virtually indistinguishable [see ESI].) 
As we have found
that $n_{\rm cl}^\ddagger$ remains roughly constant for $W\gtrsim
3.5$\,nm, the remaining source for deviations of the rate from that of
bulk may lie in the dynamics, as codified by $f^+$. In
Fig.~\ref{fig:Seeding_Figures}c we present the attachment frequency
obtained from
\begin{equation}
  f^+ = \frac{\big\langle\big(n_{\rm cl}(t)-n_{\rm cl}^\ddagger\big)^2\big\rangle}{2t},
\end{equation}
as first proposed by Auer and Frenkel.\cite{Auer2001,Auer2004} In
practice, $f^{+}$ is obtained by starting many simulations with
$n_{\rm cl}(t=0)=n_{\rm cl}^\ddagger$ and fitting
$\big\langle\big(n_{\rm cl}(t)-n_{\rm cl}^\ddagger\big)^2\big\rangle$
to a straight line after an initial transient time. Despite
significant noise in $\big\langle\big(n_{\rm cl}(t)-n_{\rm
  cl}^\ddagger\big)^2\big\rangle$, it is clear that the attachment
frequency is largely insensitive to $W$. Bringing together the above,
in Fig.~\ref{fig:Seeding_Figures}d we present $J(W)$ for the different
film thicknesses investigated (see also
Table~\ref{tab:Results_Table}). As a check of our implementation, our
result for $J(\infty)$---obtained from a simulation of bulk water
under periodic boundary conditions---agrees well with that previously
reported by Sanchez-Burgos \etal{}\cite{Sanchez2022} at $T=220$\,K and
$p=1$\,bar. The biggest drawbacks of the seeding method are that it
presupposes the nucleation mechanism and relies upon CNT to compute
$J$. \tcb{Previously, Lupi \etal{}\cite{lupi2017role} have provided 
strong evidence that using the size of the critical cluster is a 
good ``reaction coordinate'' for nucleation; in the ESI we 
present committor analyses for $W = 5$\,nm and 
$W = 2.5$\,nm that suggests that this is also the case for the thin film 
systems we consider.}

\subsection{Comparing, and reconciling, results from seeding with previous studies}

Despite good agreement of $J(\infty)$ with previous literature values,
our results for $W\sim 5$\,nm are at odds with previous work. For
example, using forward flux sampling, for mW under the same
conditions, Haji-Akbari \etal{}\cite{HajiAkbari2014} found $J(5\,{\rm
  nm})/J(\infty)\approx 0.006$. Similarly, by computing mean first
passage times at 205\,K, L\"{u} \etal{}\cite{Lu2013} found that ice
nucleation rates in thin water films were suppressed compared to that
of bulk. How can we rationalize this apparent discrepancy of our
seeding simulations, which find that $J(W)/J(\infty)\approx 1$ once
the film is thick enough to accommodate a critical nucleus? We, in
fact, argue that this discrepancy is largely superficial. In the case
of L\"{u} \etal, the reported impact of $W$ on the rate was relatively
modest, with $J(5\,{\rm nm})/J(\infty) \approx 0.5$, within the range of
uncertainty of our result. But there is also a more subtle aspect at
play. Methods such as forward flux sampling and calculating the mean
first passage time obtain a rate first by computing the probability to
undergo a nucleation event per unit time in a particular sample, and
then normalize by the volume of the sample. For homogeneous systems,
the definition of this volume is unambiguous. In contrast, for
inhomogeneous systems such as thin water films, what to take for the
volume is less clear-cut, and $J(W)$ becomes increasingly sensitive to
this normalization procedure as $W$ decreases. The seeding method, on
the other hand, provides an estimate of the rate directly (see
Eq.~\ref{eqn:JCNT}), albeit conditioned, in our study, on the nucleus
forming in the bulk-like region. The advantage of the seeding method
is that it clearly demonstrates that $n_{\rm cl}^\ddagger$ and $f^+$
remain roughly constant for $W\ge 3.5$\,nm. We note that in the
supporting information of Li \etal,\cite{Li2013} who used forward flux
sampling, the nucleation rate in a film comprising 4096 mW water
molecules was found to be indistinguishable from that of bulk; it
appears that Li \etal{} account for the surface region in their
normalization procedure.

\begin{figure*}[t]
  \centering
  \includegraphics[width=16cm]{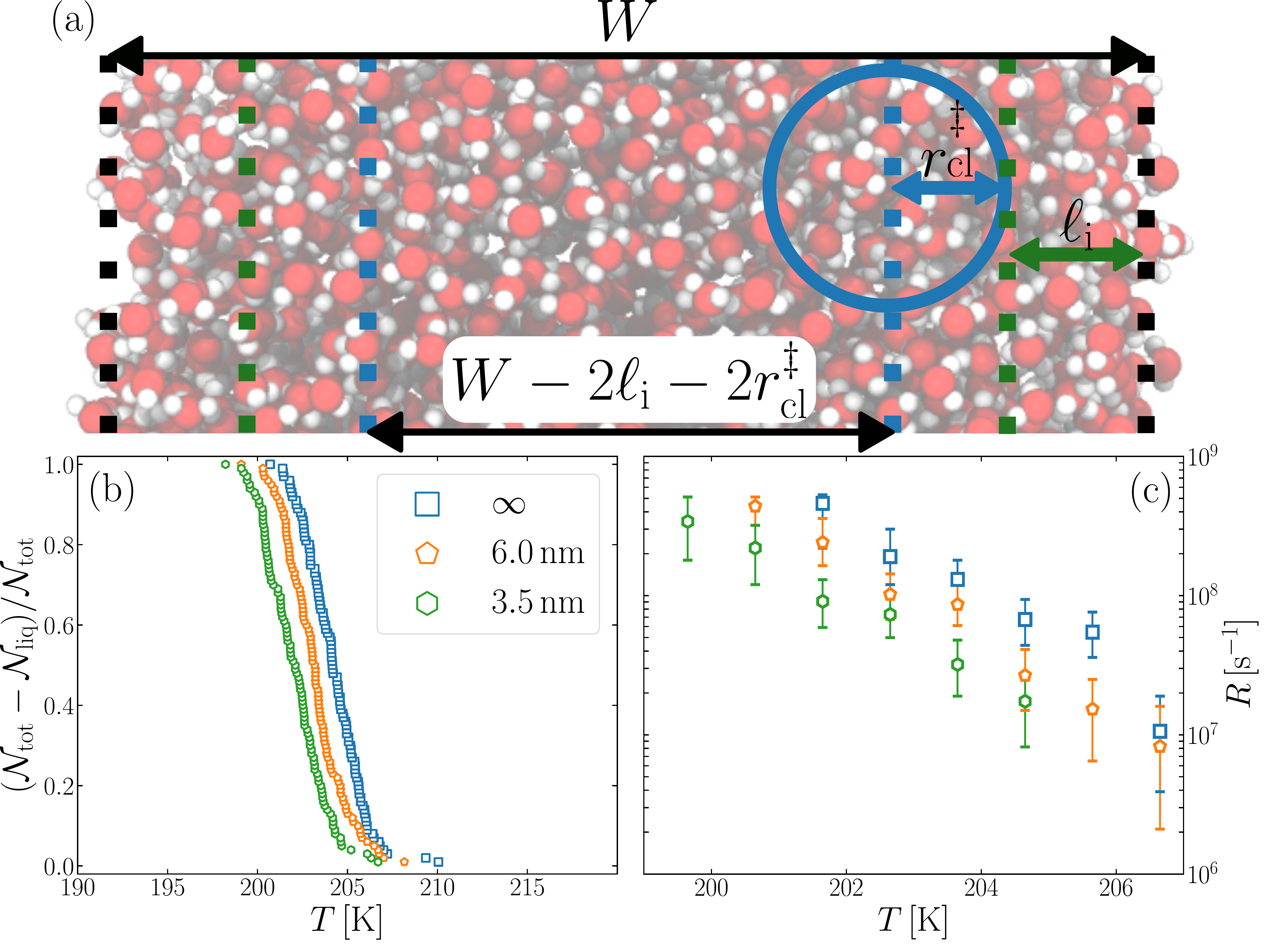}
  \caption{Reconciling nucleation rates obtained from seeding with
    other approaches. (a) Schematic of the simple model described in
    the text. Nucleation is assumed to proceed in a bulk-like fashion
    in the central volume $A(W-2\ell_{\rm i}-2r_{\rm
      cl}^\ddagger)$. In (b) we present the fraction of frozen samples
    for $W/{\rm nm}=3.5$ and $6$, and bulk water. Each system
    comprises 6000 mW molecules and is cooled at a rate of 0.2\,K/ns
    from an initial temperature of 220\,K. There is a modest shift to
    lower temperatures with the films compared to bulk water. The
    freezing rate, $R$, for $W=3.5$\,nm, as shown in (c), differs at
    most by a factor five compared to the bulk result.}
    \label{fig:Cooling_Ramps}
\end{figure*}

To help further understand the differences between $J(W)$ computed
with seeding and previous works, it is instructive to introduce the
following simple model, which is similar to existing models introduced
by others.\cite{Lu2013,Li2013,HajiAkbariPerspective2017} We suppose
that the thin film can be separated into two interfacial regions of
thickness $\ell_{\rm i}$ that sandwich a central bulk-like region of
thickness $W-2\ell_{\rm i}$. Accounting for the volume occupied by the
critical nucleus, the volume accessible for nucleation in the
bulk-like region is $A(W-2\ell_{\rm i}-2r_{\rm cl}^\ddagger)$ (see
Fig.~\ref{fig:Cooling_Ramps}). In the remaining volume, $2A(\ell_{\rm
  i}+r_{\rm cl}^\ddagger)$, we assume that nucleation occurs with a
rate $J_{\rm i}$. For a sample of total volume $AW$, the number of
nuclei that form per unit time is determined by an effective
nucleation rate $J_{\rm eff}$:
\begin{equation}
  AWJ_{\rm eff}(W) = 2A(\ell_{\rm i}+r_{\rm cl}^\ddagger)J_{\rm i} + A(W-2\ell_{\rm i}-2r_{\rm cl}^\ddagger)J(\infty).
  \label{eqn:Jeff1}
\end{equation}
Rearranging for $J_{\rm eff}$, and assuming that $J_{\rm i} \approx
0$, we find
\begin{equation}
  J_{\rm eff}(W) = \bigg[1 - \frac{2(\ell_{\rm i}+r_{\rm cl}^\ddagger)}{W}\bigg] J(\infty).
  \label{eqn:Jeff2}
\end{equation}
Using $r_{\rm cl}^\ddagger \approx 0.96$\,nm obtained from our seeding
simulations, a suppression of the nucleation rate of $J_{\rm
  eff}(5\,\mathrm{nm})/J(\infty) \approx 0.006$, as obtained from forward flux
sampling,\cite{HajiAkbari2014} requires $\ell_{\rm i} \approx
1.53$\,nm.  While this order of magnitude is reasonable for an
interfacial thickness, it is inconsistent with our finding that the
size of critical nucleus remains constant for $W\gtrsim 3.5$\,nm. As
we did above, we can instead estimate $\ell_{\rm i} = (3.5\,{\rm
  nm}-2r_{\rm cl}^\ddagger)/2 \approx 0.8$\,nm. For $W=5$\,nm, we then
obtain $J_{\rm eff}(5\,\mathrm{nm})/J(\infty) \approx 0.3$. This more modest
suppression in the rate appears comparable to that obtained at 205\,K
by L\"{u} \etal\cite{Lu2013}

\subsection{Probing the effective rate in thin water films through the lens of an experimentalist}

What the above analysis demonstrates is the sensitivity of the
effective nucleation rate to the interfacial thickness $\ell_{\rm i}$
and the size of the critical nucleus $r_{\rm cl}^\ddagger$. While the
latter is in principle a physical observable, there is a degree of
arbitrariness in defining an interfacial thickness, which obfuscates
direct interpretation of $J_{\rm eff}$. In an attempt to gauge the
significance of any suppression of $J_{\rm eff}$ with $W$, it is
useful to think of an instance when one might be interested in ice
formation in such small volumes, such as to understand how a porous
material (or a rough surface with, e.g., cracks) might promote ice
formation via a pore condensation and freezing
mechanism.\cite{David2019, Holden2021,Campbell2018,Whale2017,Pach2019}
In such cases, one is probably less interested in the precise value of
an effective rate, and instead more concerned whether ice can form,
given that a pore/crack is filled with water under the conditions.

In this spirit, in Fig.~\ref{fig:Cooling_Ramps}b we present the
fraction of frozen samples against temperature for bulk water,
$W=6\,\mathrm{nm}$ and $W=3.5\,\mathrm{nm}$, obtained by cooling 100
replicas of each system at $0.2\,\mathrm{K\,ns^{-1}}$. Note that the
number of water molecules is the same in each case (6000 mW). The size
of the largest ice-like cluster was monitored, and the system was
considered frozen when the largest cluster $n_\mathrm{cl}(T) \geq
200$.
\footnote{In rare cases that the largest nucleus recrossed 200
molecules, the lower temperature for which $n_\mathrm{cl}(T)$ passed
200 molecules was taken. Such recrossings occurred in fewer than 1\%
of our simulations. } We see that the curves are shifted relative to
each other, with the bulk samples freezing at the highest temperature,
and the $W=3.5\,\mathrm{nm}$ film at the lowest temperature being
separated by around 2\,K. We are able to achieve statistically
different \tcb{freezing temperatures and rates} in our studies
primarily due the artificially high control we can exert on the number
of molecules ($N = 6000$) and with it the volume undergoing
cooling.

In experimental systems such fine control is impossible, and with it
the deviations in freezing temperatures and freezing rates reported
here become unimportant. To see this, we compare the impact of varying
$W$ by computing the freezing rate, $R$, for each of the data sets
using an approach typically employed in the analysis of laboratory
experiments.\cite{vali2015proposal} $R$ is analogous to the radioactive decay constant (a
first-order reaction rate constant) and is most easily determined from
experiments conducted on completely identical supercooled droplets
under isothermal conditions. In the isothermal case, the fraction of
droplets which freeze in a given time period can be found
as:\cite{koop1997freezing}
\begin{equation}
  \frac{\mathcal{N}_{\mathrm{liq}}(t)}{\mathcal{N}_{\rm tot}}=\exp (-Rt)
  \label{eqn:Surviving_Droplets}
\end{equation}
where $\mathcal{N}_{\mathrm{liq}}(t)$ is the number of liquid droplets
remaining after time $t$ and $\mathcal{N}_{\rm tot}$ is the total
number of droplets (liquid or solid) present in the experiment [equal
  to $\mathcal{N}_{\mathrm{liq}}(t=0)$]. In order to calculate $R$
from continuously cooled experiments, it is necessary to divide the
experiment into small time intervals $\Delta t$ over which changes in
temperature are treated as small. Rearranging
Eqn.~\ref{eqn:Surviving_Droplets} we find
\begin{equation}
    R=\frac{-\log \left(1-\mathcal{N}_{\mathrm{f}} / \mathcal{N}_{\mathrm{i}}\right)}{\Delta t}
\end{equation}
where $\mathcal{N}_{\mathrm{i}}$ is the number of unfrozen droplets
present at the start of the time interval and $\mathcal{N}_\mathrm{f}$
is the number of droplets which freeze during the time interval. We
have calculated $R$ for the three systems by treating each of their
100 replicas as a droplet. We have used $\Delta t = 5\times
10^{-9}$\,s, resulting in temperature bins 1\,K wide.  The results of
this analysis are shown in Fig.~\ref{fig:Cooling_Ramps}c, where the
data points indicate the midpoint of the temperature bins.  Confidence
intervals were calculated using a simple Monte Carlo simulation
implemented in Stata 17.\cite{Stata} For physically identical droplets
it is expected that the number of freezing events in a temperature
interval will follow a Poisson distribution on repeat
testing.\cite{koop1997freezing,blackman2017permeable} For each
temperature bin, the observed number of freezing events was taken as
the expectation value for the bin. Poisson distributed random numbers
were then generated to create 5000 simulated repeats of each of the
three data sets. The bars on Fig. \ref{fig:Cooling_Ramps}c show the
5$^{\mathrm{th}}$ to 95$^{\mathrm{th}}$ percentile range of the
simulated experiments. (Data points generated from bins containing so
few freezing events that the 5$^{\mathrm{th}}$ percentile simulation
had a value of $R=0$ have not been included in
Fig.~\ref{fig:Cooling_Ramps}c.) The freezing rates determined for the
bulk system are clearly higher than those found for
$W=3.5\,\mathrm{nm}$. This difference is largest for the bins centered
at 201.65\,K, where the bulk rate is greater by approx. a factor of
five.

In principle, the homogeneous nucleation rate, $J$, can be found as
$R/V$ where $V$ is the volume of the water droplets undergoing
freezing. Taking the volume of an mW water molecule in the liquid
phase as $1/\bar{\rho} \approx 3 \times 10^{-29}\,\mathrm{m^{3}}$ we
find that 6000 water molecules occupy $1.8 \times
10^{-25}\,\mathrm{m^{3}}$. Dividing our calculated freezing rates by
this volume we obtain values of $J$ entirely consistent with previous
literature values
\cite{HajiAkbari2014,russo2014new,moore2011structural,
  Sanchez2022} for mW (see ESI). However,
obtaining $J$ in this fashion is made possible as we know the exact
number of water molecules used in our simulations. Experimentally, it
is infeasible to determine the volume with such precision, so the
differences in rates we report are likely inconsequential in any
practical setting.

%% produce the $J$ values shown in the ESI and compared to
%% literature data. It can be seen that the nucleation rate produced from
%% the cooling approach used here generates results entirely compatible
%% with previous calculations of the homogeneous ice nucleation using the
%% mW model. We note also that the bulk nucleation rate calculated later
%% for 6000 mW water molecules at $220\,$K agrees very well with previous
%% calculations. 

\section{Assessing the impact of physical boundaries}
\label{sec:PhysicalBoundaries}

\begin{figure*}[t]
  \centering \includegraphics[width=16cm]
             {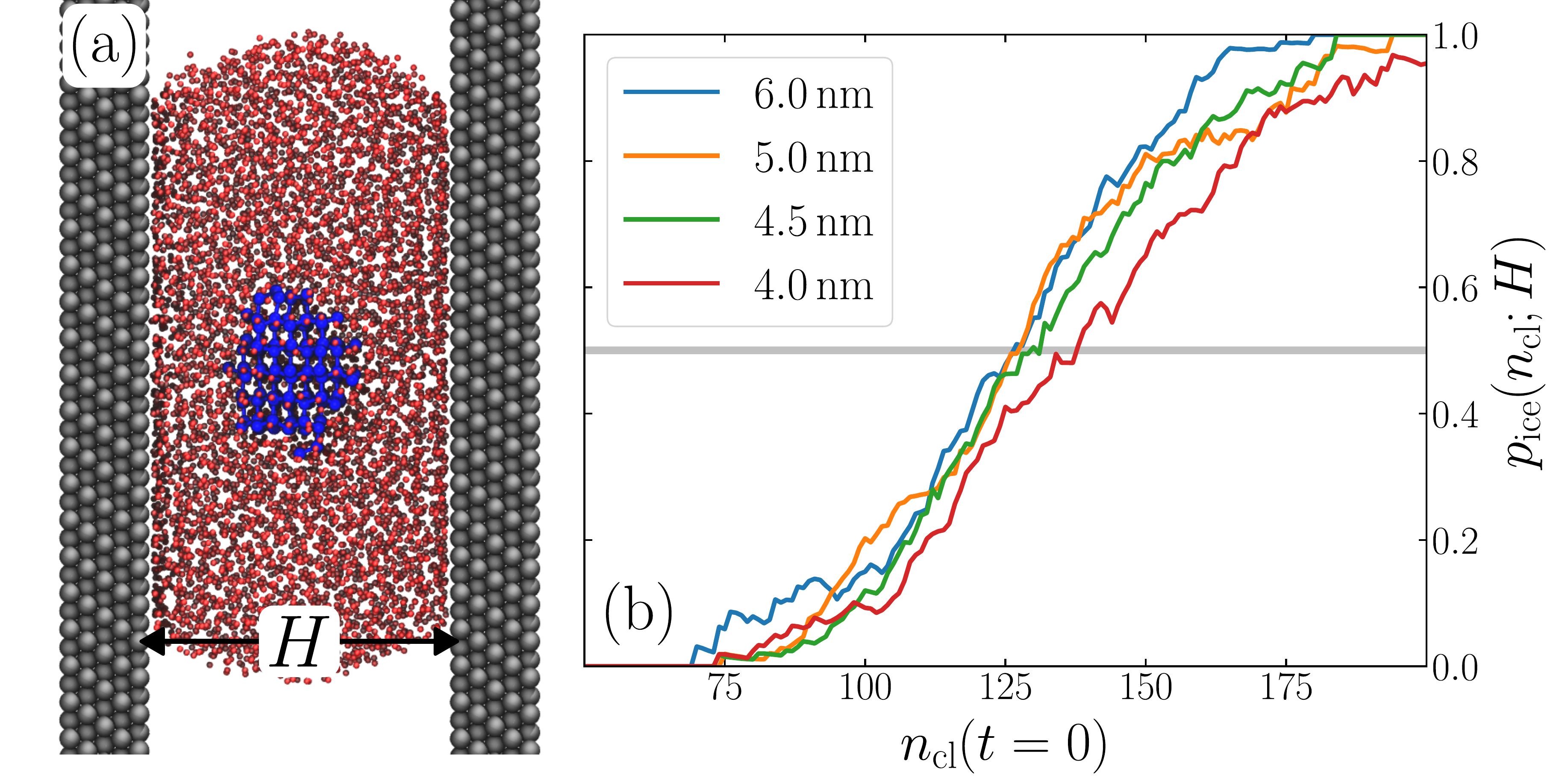}
  \caption{Conclusions drawn from unsupported thin films appear robust
    to the presence of physical boundaries. (a) Snapshot from a seeding
    simulation with a supported film confined between two slabs of
    Lennard-Jones particles separated by $H = 4.5$\,nm. The water
    forms a `squished cylinder' that spans the periodic boundaries of
    the simulation cell out of the plane of the page. In panel (b), we
    show $p_{\mathrm{ice}}(n_{\mathrm{cl}};H)$, which shows the
    $n_{\rm cl}^\ddagger$ remains bulk-like for $H\gtrsim
    4.5$\,nm. For $H = 4$\,nm, the size of the critical nucleus
    increases, similar to what we observed in the unsupported films.}
    \label{fig:Heterogeneous_Figures}
\end{figure*}

The thin water films we have investigated so far are able to provide
insight into the fundamental question of how much liquid water is
needed to recover bulk-like nucleation. But, the direct relevance of
thin water films in coexistence with vapor to real physical scenarios
is, in fact, somewhat limited; small volumes of liquid water will form
droplets with high interfacial curvature, with an associated Laplace
pressure. As discussed in Sec.~\ref{sec:Intro}, Li \etal\cite{Li2013}
have found that differences in ice nucleation rate in small droplets
relative to bulk water can largely be accounted for by this Laplace
pressure. Notwithstanding the issues discussed in
Sec.~\ref{sec:Thin_Films} in calculating the rate in small systems, we
argue that this analysis of Li \etal{} is in line with our finding
that $J(W)/J(\infty) \approx 1$ once the film is thick enough to
accommodate a critical nucleus. Instead of extending our studies to
small droplets, then, we instead focus on a system whose direct
comparison to thin water films is more apparent: water confined
between two `inactive' walls.

A snapshot of the system we simulate is shown in
Fig.~\ref{fig:Heterogeneous_Figures}a. To introduce confining walls, a
slab comprising 9600 atoms arranged on a FCC lattice, and exposing its
(111) face, is placed such that its lowermost plane of atoms is
situated at $z = H/2$. An identical slab is then placed such that its
uppermost plane of atoms is located at $z=-H/2$. The atoms belonging
to these slabs interact with mW water molecules by a Lennard-Jones
potential. The lattice constant of the slab and the parameters for the
Lennard-Jones potential (see ESI) were chosen such that the
surface only promotes ice formation at temperatures below
$201\,\mathrm{K}$.\cite{Davies2022} As we are investigating nucleation
at $T=220$\,K, we can consider these surfaces to be inactive to ice
nucleation, and we will assume that nucleation occurs homogeneously in
the interior `bulk-like' region of the fluid. Between these slabs, we
introduce mW water molecules such that a `squished cylinder' forms
that spans one of the lateral dimensions, shown in
Fig.~\ref{fig:Heterogeneous_Figures}a. In the orthogonal lateral
dimension, a slightly convex water-vapor interface forms. The presence
of a curved interface implies that, unlike the thin films considered
above, the pressure inside the fluid is nonzero. In the films
considered, the density far from the interfaces is virtually
indistinguishable from the case of the thin water film (see
ESI).  Note that the amount of water included in the simulation
depends on $H$ such that the curvature of the liquid-vapor interface
is approximately constant (see ESI).

\begin{figure}[h]
  \centering
  \includegraphics[width=7.5cm]
  {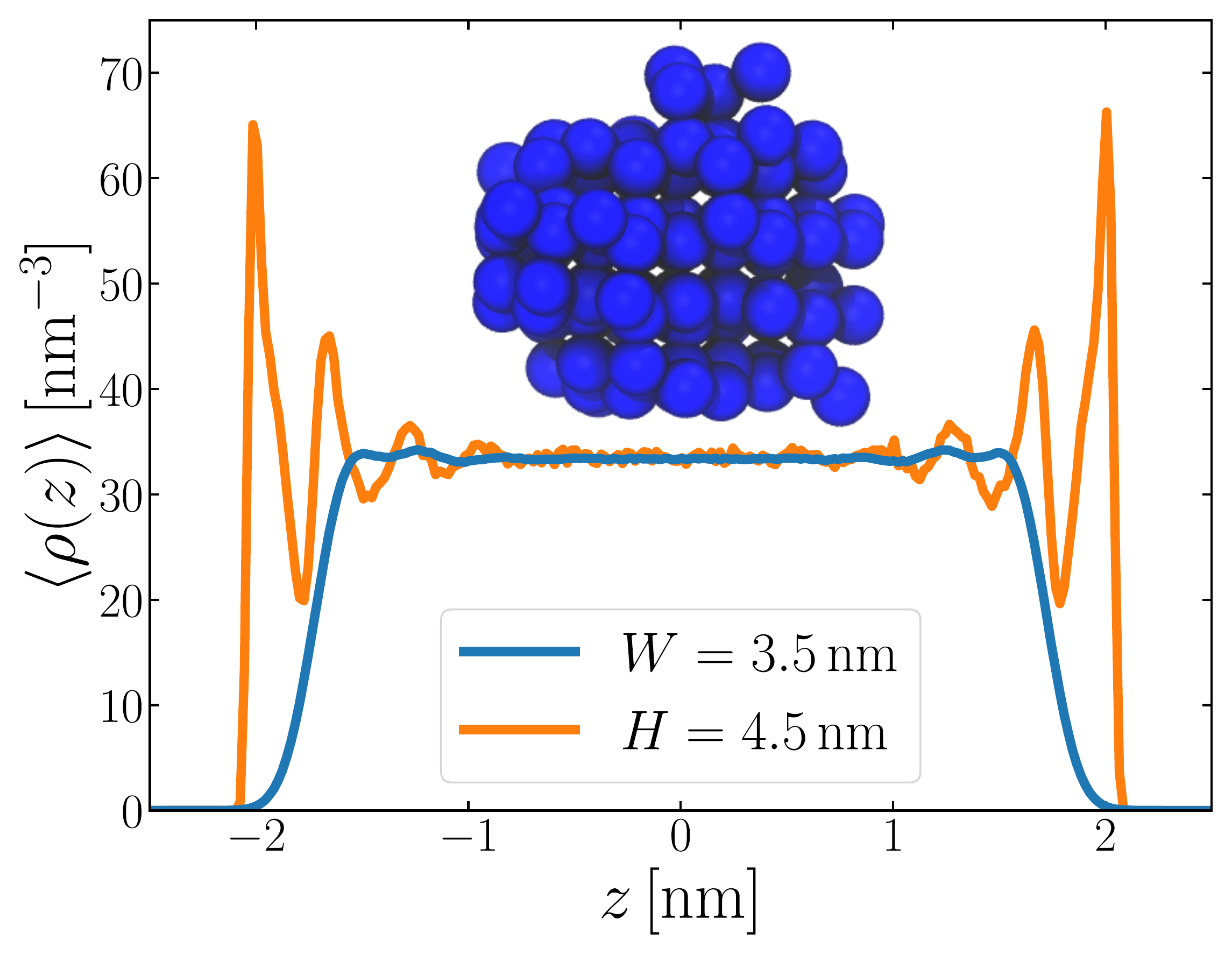}
  \caption{The smallest values of $W$ (unsupported film) and $H$
    (confined between walls) for which $n_{\rm cl}^\ddagger$ remains
    bulk-like both exhibit similar sized bulk-like regions, as seen in
    $\langle\rho(z)\rangle$. For the system confined between walls,
    $\langle\rho(z)\rangle$ is calculated such that we only consider a
    region far from the convex liquid/vapor interface. Overlayed on these
    plots is a snapshot of a critical nucleus, roughly to scale, which
    shows that it spans essentially all of the central bulk-like
    region.}
    \label{fig:Density_Profiles}
\end{figure}

Following the same procedure as we did in Sec.~\ref{sec:Thin_Films},
we performed seeding simulations for $H/{\rm nm}=4.0, 4.5, 5.0$ and
$6.0$. In Fig.~\ref{fig:Heterogeneous_Figures}, we present $p_{\rm
  ice}(n_{\rm cl}; H)$, which shows that for $H\ge 4.5$\,nm, the size
of the critical nucleus is insensitive to $H$ and, within statistical
uncertainty, indiscernible from $n_{\rm cl}^\ddagger$ found in the
thicker unsupported thin films. Also similar to the case of the
unsupported films, for the smallest value of $H$ investigated, $n_{\rm
  cl}^\ddagger$ is seen to increase. Providing a rigorous relationship
between $H$ and $W$ is beyond the scope of this work. Instead, in
Fig.~\ref{fig:Density_Profiles} we present density profiles of water
for both the $H=4.5$\,nm confined system, and the $W = 3.5$\,nm
unsupported film (i.e., the smallest value of $H$ and $W$ for which
$n_{\rm cl}^\ddagger$ remains bulk-like). By eye, the extent of the
bulk-like region in these two systems is comparable. Superposed on
these plots is a snapshot (roughly to scale) of a critical nucleus for
the $W=3.5$\,nm system, which shows that the diameter of the critical
nucleus occupies the full extent of this bulk-like region. The insight
gleaned from our investigation on thin films therefore also appears to
hold in the more realistic case that water is confined between
surfaces that do not promote ice formation.

\section{Conclusions}
\label{sec:Concl}

In this article, we have investigated ice nucleation in thin films of
water, both freestanding, and confined between surfaces that do not
promote ice nucleation. In both cases, our results show that the
critical nucleus' size is indistinguishable from that of bulk water
for sample sizes that can barely accommodate its presence. At a
temperature of 220\,K, once the thickness is decreased below
approx. 3.5\,nm, we found that the size of the critical nucleus
increases. These results for thin films are consistent with the
observation in Ref.~\citenum{Li2013} that bulk-like nucleation is
recovered in nanometer sized droplets, once the Laplace pressure is
taken into account, and indeed, that the rate was seen to decrease
relative to that of bulk for droplets too small to accommodate a
critical nucleus. Ref.~\citenum{Li2013} and this work, however, have
reached this conclusion using a coarse grained representation of
water's interactions (the mW model), while previous
work\cite{HajiAkbari2017} has found a qualitative discrepancy with
more a more detailed model (TIP4P/ice) when ice nucleation rates in
thin films are compared to those in homogeneous bulk. So, in this
work, we have also shown that this apparent discrepancy between water
models can be resolved through consistent treatment of truncated
interactions between the homogeneous and inhomogeneous systems, when
using more detailed models such as TIP4P/ice.

In the thin film systems we have investigated, we also observed that
the attachment frequencies to the critical nuclei are comparable to
those in bulk water. Calculating the rate based on classical
nucleation theory then gives the impression that the nucleation rate
in thin films is the same as bulk water, which is seemingly at odds
with some,\cite{Lu2013,HajiAkbari2014,HajiAkbari2017,Vrbka2006} but
not all,\cite{Li2013} previous works. A simple model that supposes
nucleation is suppressed in the interfacial regions is able to
reconcile these differences at a qualitative level, but drawing
quantitative conclusions on the rate (per unit volume per unit time)
is made challenging by the sensitivity of the effective rate to the
definition of the interfacial thickness. One scenario where one might
be interested in the kinetics of ice formation under confinement is in
the presence of pores or microstructures presented by solid particles
such as silicas or feldspars i.e., pore condensation and
freezing.\cite{Holden2021} Describing ice formation in these complex
systems may benefit from simplified models,
\tcb{and to this end, the results of this
work suggest CNT provides a reasonable foundation.}  

%% well-parameterized \tcb{models that apply CNT to more 
%% complex heterogeneous systems could be
%% used to good effect (see e.g., Refs.~\citenum{Rosky2023}
%% and~\citenum{David2019}).}

%Describing ice formation in these complex
%systems may benefit from simplified models, and the results of this
%work suggest well-parameterized classical nucleation theories could be
%used to good effect (see e.g., Refs.~\citenum{Rosky2023}
%and~\citenum{David2019}).}

In any case, in experiments that investigate pore condensation and
freezing, one typically measures the temperature at which ice forms
rather than the nucleation rate directly. To gauge the impact on
readily obtainable experimental observables, we therefore compared the
fraction of frozen samples containing the same number of water
molecules. Our results suggest that any difference in nucleation rate
in a 3.5\,nm thick film of water relative to that of a macroscopic
sample is slight, and likely inconsequential for any experimental
measurement that can be performed in the foreseeable future. They
further suggest that the ice nucleation rate for water condensed in
the pores of aerosol particles under cirrus cloud conditions will not
be significantly suppressed due to the confined nature of the water in
those pores. 

%\tcb{By showing that these nucleation rates in pores 
%remain effectively unchanged, we provide more support for the framework 
%used by David \etal{} to show the rate determining step of PCF, 
%that of the crystal growth out of the pores, occurs at appreciable rates 
%at atmospheric conditions. Ice nucleation activity of particles in 
%cirrus clouds will therefore depend on their porosity and on the 
%wettability of their pores.}

%As such, they support the conclusion of David
%\etal\cite{David2019} that the ice nucleation activity of particles in
%cirrus clouds will depend on their porosity and on the wettability of
%their pores.

\section{Methods}\label{sec:Methods}

Full details of the methodology are given in the ESI.  Liquid
structures were initialized using the packmol software
package,\cite{Packmol} and crystal structures with
GenIce.\cite{Matsumoto:2017bk, Matsumoto:2021} All simulations were
performed with the LAMMPS simulation package\cite{LAMMPS} using the
standard velocity verlet algorithm.\cite{Swope1982} The temperature
was controlled using a Nos\'e-Hoover thermostat\cite{Nose2006} and,
where necessary, the pressure maintained with a 
Parinello-Rahman barostat.\cite{parrinello1981polymorphic} 
Seeding systems were produced with the aid of the MDAnalysis
software package.\cite{Michaud2011,beckstein2016}

For TIP4P/ice,\cite{Abascal2005} simulations were performed in a
similar manner, with electrostatic interactions computed with a
particle-particle particle-mesh Ewald
method,\cite{hockney2021computer} with parameters chosen such that the
root mean square error in the forces were a factor $10^5$ smaller than
the force between two unit charges separated by a distance of
1\,\AA.\cite{Kolafa1992} For simulations of liquid water in contact
with its vapor, we set $D=0$, where $D$ is the electric displacement
field along $z$, using the implementation described in
Refs.~\citenum{Sayer2019}~and~\citenum{Cox2019} (this is formally equivalent
to the commonly used Yeh and Berkowitz\cite{Yeh1999} method). The
rigid geometry of the molecules was maintained with the RATTLE
algorithm.\cite{Anderson1983} Parallel
tempering\cite{swendsen1986replica, earl2005parallel} was employed for
the TIP4P/ice$^{(8.5)}$ film in Sec.~\ref{subsec:Justifying_mW}.

Steinhardt order parameters were calculated using the 
open-source, community-developed PLUMED library,\cite{Bonomi2019}
version 2.8.\cite{Tribello2014} Snapshots were created in
VMD.\cite{HUMP96, Ston1998}

\section*{Data Availability Statement}

The data that supports the findings of this study and input files for
the simulations, are openly 
available at the University of Cambridge
Data Repository, \url{https://doi.org/10.17863/CAM.96642}.

\section*{Author Contributions}

\textbf{John A. Hayton}: Data Curation (lead), Formal Analysis (lead),
Investigation (lead), Methodology (lead), Writing - Original Draft
(lead), Writing - Review and Editing (contributing)

\textbf{Michael B. Davies}: Conceptualization (contributing), Supervision (contributing), Writing - Review and Editing (contributing)

\textbf{Thomas F. Whale}: Formal Analysis (contributing), Writing - Review and Editing (contributing)

\textbf{Angelos Michaelides}: Conceptualization (contributing),  Writing - Review and Editing (contributing)

\textbf{Stephen J. Cox}: Conceptualization (lead), Supervision (lead), Investigation (contributing)
Writing - Review and Editing (lead)

\section*{Conflicts of interest}

There are no conflicts to declare.

\section*{Acknowledgements}

J.A.H. acknowledges studentship funding from the Engineering and
Physical Sciences Research Council (EP/T517847/1). 
T.F.W. thanks the Leverhulme Trust
and the University of Warwick for supporting an Early Career
Fellowship (ECF-2018-127). S.J.C. is a Royal Society University
Research Fellow (Grant No. URF\textbackslash R1\textbackslash 211144)
at the University of Cambridge.

\renewcommand\refname{References}

%%%REFERENCES%%%
\bibliography{Sources} %You need to replace "rsc" on this line with the name of your .bib file
\bibliographystyle{rsc} %the RSC's .bst file

\newpage

\onecolumngrid

\renewcommand\thefigure{S\arabic{figure}}
\renewcommand\theequation{S\arabic{equation}}
\renewcommand\thetable{S\arabic{table}}
\renewcommand\thesection{S\arabic{section}}
\setcounter{figure}{0}
\setcounter{equation}{0}
\setcounter{table}{0}
\setcounter{section}{0}

\begin{center}
    {\Large Electronic Supporting Information for The limit of macroscopic 
homogeneous ice nucleation at the nanoscale}
\end{center}

\maketitle

  This document includes: methodology for all simulations in the main
  paper; a sensitivity analysis of the nucleation rate to $\Delta\mu$;
  committor distributions for 2.5\,nm and 5\,nm films; density
  profiles for the physically confined films; volumetric nucleation
  rates obtained by seeding and cooling ramps compared to literature
  values; and preliminary work using seeding comparing bulk to film
  samples in TIP4P/ice$^{(8.5)}$.

\section{Methodology}\label{subsec:Methodology_SI}

All simulations were performed with LAMMPS,\cite{LAMMPS} with particle
positions propagated with the velocity verlet algorithm.\cite{Swope1982} The timestep,
$\delta t$, was set at $10\,\mathrm{fs}$ in mW simulations and
$2\,\mathrm{fs}$ in TIP4P/ice$^{(8.5)}$ simulations. Simulations of
thin water films employed a slab geometry under periodic boundary
conditions, with the temperature maintained with a Nos\'e-Hoover
thermostat with a damping constant $100\delta t$.  Simulations of bulk
water were performed under periodic conditions in the NPT
ensemble; temperature was similarly maintained with a Nos\'{e}-Hoover
thermostat\cite{Nose2006}, and pressure maintained at 0\,bar with a
Parinello-Rahman barostat\cite{parrinello1981polymorphic} with 
damping constant $1000\delta t$.

\subsection{Preparation of thin mW films}\label{subsec:Prep_mW_Films}

For unsupported thin films, the thickness $W$ was defined as
\begin{equation}
    W = N / A \bar{\rho} 
\end{equation}
with $N$ the number of molecules, $A$ the cross sectional area of the
film, and $\bar{\rho}=33.3774\,\mathrm{nm^{-3}}$ the bulk density of
mW molecules at 220\,K. By running initial tests at 6000 and
8192 molecules in the 6\,nm film, we found that there were no
significant finite size effects present for a critical seed in a 6000
particle system, akin to previous investigations that selected
similar sized lateral dimensions\cite{HajiAkbari2014,Lu2013}. $W$ was
therefore controlled by selecting the $x$ and $y$ dimensions ($l_x$
and $l_{y}$ respectively) of the film so that $l_{x} = l_y = \sqrt{N /
  W \bar{\rho}} = 13.40\,\mathrm{nm} \times \sqrt{1\,\mathrm{nm} /
  W}$.

To generate an initial configuration of mW water molecules for a given
film width, the packmol software\cite{Packmol} was used to randomly
pack 6000 mW particles in a cuboid between $(0,\,0,\,-W/2)$ and
$(l_x,\,l_y,\,W/2)$. The $z$ dimension ($l_z$) of the simulation box
was then increased by a factor of 3 to generate a liquid-vapor
interface. The films were equilibrated at 220\,K for
$10000\:\mathrm{steps}$ at timesteps of $0.1\:\mathrm{fs}$,
$0.5\:\mathrm{fs}$, $1\:\mathrm{fs}$, $2\:\mathrm{fs}$ and
$5\:\mathrm{fs}$ sequentially, before further equilibration for
$50000\:\mathrm{steps}$ at a $10\:\mathrm{fs}$ timestep.  After this,
the systems were judged to have equilibrated, as the density of the
center of the film and enthalpy had converged.

Independent initial configurations of the mW films were generated by 
performing simulations of the equilibrated film at $220\,$K and taking 
the positions of all atoms every $100\,$ps.

\subsection{Preparation of bulk mW systems}\label{subsec:Prep_mW_Bulk}

An initial random configuration of mW particles was generated using
packmol, with $l_{x} = l_y = l_{z} = (N / \bar{\rho})^{1/3}$. The
system was then equilibrated with the protocol described in
Sec.~\ref{subsec:Prep_mW_Films}, with the density and enthalpy
adjudged to have equilibrated. Independent configurations were then
sampled as described in Sec.~\ref{subsec:Prep_mW_Films}

\subsection{Preparation and simulation of mW seeding systems}
\label{subsec:mW_Seeding}

Independent configurations of mW as hexagonal ice were generated by 
first utilizing the GenIce package to create an initial configuration 
of TIP4P/ice molecules in the ice Ih crystal structure. The hydrogens 
were stripped out using python code, and the oxygens relabeled as mW 
particles, to generate a system of mW particles in the Ih crystal 
structure with $N=16000$. The system was allowed to equilibrate at 
220\,K and 0\,bar for 10\,ns.
%after which the density and enthalpy of 
%the system was adjudged to be equilibrated.
Independent configurations were then obtained by allowing the system
to further evolve at 220\,K and 0\,bar, with the positions of all
molecules taken every $100\,$ps.

To generate seeding systems, an initial configuration of mW particles
of the desired system (i.e., thin film or bulk liquid) is selected,
and python code utilizing the MDAnalysis software package\cite{Michaud2011,
beckstein2016} is used to cut a spherical cavity
%(centered at
%$(0,\,0,\,0)$, allowing for periodic boundary conditions)
in the liquid. A spherical cluster of Ih is created by cutting a
similar sphere out of an initial configuration of ice Ih. The
spherical seed is placed in the cavity cut in the liquid configuration
to initialize a seeding system.

Seeding systems progressed by first allowing the liquid molecules to
equilibrate for 80\,ps at 220\,K with the molecules in the seed held
fixed; for a seed immersed in the bulk liquid, pressure was also
maintained at 0\,bar. This was done to allow a more realistic
liquid/crystal interface to develop, with 80\,ps selected to allow
enough time for this surface to relax, but not so long that the seeds
close to the critical size grew substantially. Then, all particles
were allowed to evolve until either $n_\mathrm{cl} \gg
n_\mathrm{cl}^\ddagger$ (taken as $n_\mathrm{cl} \geq 400$) or
$n_\mathrm{cl} \approx 0$ (taken as $n_\mathrm{cl} \leq
20$). Information on how the $n_\mathrm{cl}$ was identified and
monitored is given in Sec.~\ref{subsec:Identifying_Ice}.

\subsection{Obtaining \textit{\textbf{n}}$_{\rm cl}^\ddag$ and \textit{\textbf{f}}$^{+}$ from seeding simulations}
\label{subsec:Finding_Critical_Nucleus}

The probability of growth, $p_{\rm ice}(n_{\rm cl}^{\ast}; W)$ for a
given cluster size $n_{\rm cl}^{\ast}$ is found by considering all
initial clusters of size $n_{\rm cl}^{\ast} -9 \leq n_{\rm cl}(t=0)
\leq n_{\rm cl}^{\ast} + 9$.  The probability of growth for this group
of simulations was found as $p_{\rm ice}(n_{\rm cl}^{\ast}; W) =
n(\mathrm{Grown}) / n(\mathrm{Total})$.  The bin size of 19 was
selected as clusters were regularly observed to change size by $\sim
9$ molecules between frames in analysis, and corresponded to a 5\%
change in radius between the largest and smallest extremes of the
range at criticality.  $p_{\rm ice}(n_{\rm cl}^{\ast}; W)$ was identified for all
integer $n_{\rm cl}^{\ast}$, and combined to generate Fig.~3a.
$p_{\rm ice}(n_{\rm cl}; W) = 0.5$ was then identified to find $n_{\rm
  cl}^\ddag$.  The ranges presented as in Tab.~1. were found by
splitting the entire data set (for each system) randomly into three
subsets and repeating the analysis with each subset. The largest and
smallest values obtained were quoted as the range.

Once $n_{\rm cl}^\ddag$ was identified, the attachment rate as found
by identifying all simulations for a given film width of initial
cluster size $n_{\rm cl}^{\ddag} -5 \leq n_{\rm cl}(t=0) \leq n_{\rm
  cl}^{\ddag} + 5$, and initiating simulations from these
  configurations.
%These near-critical clusters were simulated again
%post-equilibration at a rate of $4\,\mathrm{ps^{-1}}$, and the value
%$\left( n_{\rm cl}(t) - n_{\rm cl}(t=0) \right)^2$
%tracked.
  $\langle (n_{\rm cl}(t) - n_{\rm cl}^{\ddag})^2\rangle$
  was found by assuming $n_{\rm cl}(t=0) \approx n_{\rm cl}^{\ddag}$
  and taking the average from all simulations. $f^{+}$ was then found
  by applying a linear fit, and taking half the gradient.

\subsection{Preparation of TIP4P/ice systems}
\label{subsec:Prep_TIP4Pice}

All simulations of TIP4P/ice were performed in a similar manner to
their corresponding mW simulation, with electrostatic interactions
computed with a particle-particle particle-mesh Ewald 
method,\cite{hockney2021computer} with
parameters chosen such that the root mean square error in the forces
were a factor $10^5$ smaller than the force between two unit charges
separated by a distance of 1\,\AA.\cite{Kolafa1992} For
simulations of liquid water in contact with its vapor, we set the
electric displacement field (along $z$) $D=0$, using the
implementation given in Refs.~\citenum{Sayer2019,Cox2019}. The rigid
geometry of the molecules was maintained with the RATTLE algorithm
\cite{Anderson1983}.

TIP4P/ice systems were generated using a similar methodology to mW.
The only film generated was made to match the film considered in
Ref.~\citenum{HajiAkbari2017}, which had $N=3072$ and $l_{x} = l_{y} =
4.8096\,\mathrm{nm}$. Given the bulk density of TIP4P/ice$^{(8.5)}$ at
230\,K was measured in this study as $\bar{\rho}=30.9366\,
\mathrm{nm^{-3}}$, this gives $W=4.3\,\mathrm{nm}$. Similar to mW, the
film was generated by creating an initial random configuration of
TIP4P/ice molecules with packmol, before allowing the system to
equilibrate. The bulk system was also generated to match
Ref.~\citenum{HajiAkbari2017}, with $N=4096$. It was generated and
equilibrated in a manner similar to Sec.~\ref{subsec:Prep_mW_Bulk},
but with the relevant fixes described above. In addition, to
facilitate equilibration at this low temperature, we used replica
exchange with temperature 230, 236, 242, 248 and 254\,K; swaps between
replicas were attempted every
1\,ps.\cite{swendsen1986replica,earl2005parallel}

%% Additionally, for the profile of TIP4P/ice$^{(8.5)}$ presented, the
%% properties of the film were probed using parallel tempering.  Replicas
%% of the film were simulated at 230, 236, 242, 248 and 254\,K, with
%% swaps attempted between a random pair of replicas every
%% 1\,ps. \tcr{[** So, the profile we show was obtained with parallel
%%     tempering?]}

\subsection{Simulations with cooling ramps}

Cooling ramp simulations were performed for $W=\infty 
,\,6.0\,\mathrm{nm} , \, 3.5\,\mathrm{nm}$. The initial configurations 
generated in Secs.~\ref{subsec:Prep_mW_Films}, 
\ref{subsec:Prep_mW_Bulk} were 
taken, and were cooled at a rate of $0.2\,\mathrm{K\,ns^{-1}}$ using 
a Nos\'e-Hoover thermostat with damping parameter 1000\,fs. The 
systems were monitored every 100\,ps, and the system was adjudged to be 
frozen when $n_\mathrm{cl} \geq 200$. In the rare case that a system 
recrossed 200\,molecules, the later temperature at which 
$n_\mathrm{cl}$ passed 200 molecules was taken as the freezing 
temperature. Information on how the $n_\mathrm{cl}$ was identified and 
monitored is given in Sec.~\ref{subsec:Identifying_Ice}. 

\subsection{Preparation of systems with physical confining boundaries}

mW films were confined between the (111) face of a Lennard-Jones
crystal, with parameters described in
Tab.~\ref{tab:Het_Nuc_Parameters}. This surface had been previously
investigate by Davies \etal{}\cite{Davies2022}, and shown to nucleate
ice only for temperatures $T < 201\,\mathrm{K}$, significantly below
the temperature of 220\,K used for the seeding investigations
performed in this study.

\begin{table}[h]
    \centering
    \begin{tabular*}{0.5\textwidth}{@{\extracolsep{\fill}}ll}
        \hline
         Parameter & Value \\
         \hline
         $\epsilon$ & 0.23 $\mathrm{kcal\,mol^{-1}}$\\
         $\sigma$ & 2.5233 $\text{\AA}$\\
         LJ Cutoff & 7.53 $\text{\AA}$\\
         Lattice Constant & 4.1564 $\text{\AA}$\\
         \hline
    \end{tabular*}
    \caption{Parameters for the LJ crystal used to confine the thin 
    film. This crystal face is shown to only nucleate ice below 201\,K\cite{Davies2022}, and so we consider it nucleation inactive for this study, with any nucleation events occurring at 220\,K expected to happen far from the interface}
    \label{tab:Het_Nuc_Parameters}
\end{table}

1920 Lennard-Jones atoms were packed into a close-packed layer, and 5
such layers were combined to generate an FCC crystal with its
close-packed face exposed, perpendicular to the $z$ axis.  This
resulted in a crystal structure with dimensions $5.5778\,\mathrm{nm}
\times 24.4336\,\mathrm{nm} \times 1.1950\,\mathrm{nm}$. This crystal
was placed in a simulation cell with its uppermost layer at a position
$z=-H/2$. An identical structure was placed with its lowermost layer
at $z=H/2$.

Between these two layers, packmol was used to generate a random
configuration of $N(H)$ mW molecules in a cuboid between $(0,\, 0,\,
-H/2)$ and $(5.5778\,\mathrm{nm}, \, 8\,\mathrm{nm}, \, H/2)$.
$N(H)=1570 \times \frac{H}{[1\,\mathrm{nm}]}$ was selected so that the
physically confined film spanned the $x$ dimension of the LJ crystal,
and approximately $1/3$ of the $y$. 

%This left a large portion of the physically confined film suitably far
%from the liquid-vapour interface (forming perpendicular to the y axis)
%for nucleation behavior to be not noticeably affected by this surface.

The system was then allowed to equilibrate, and independent initial 
configurations sampled, in a manner similar to 
Sec.~\ref{subsec:Prep_mW_Films}. Seeding systems were the generated and 
investigated in a manner similar to Sec.~\ref{subsec:mW_Seeding}.
The critical nucleus size was identified in a manner identical to 
Sec.~\ref{subsec:Finding_Critical_Nucleus}.

The particles that make up the LJ crystal were not included in the 
dynamics, and did not contribute towards the simulation beyond their 
interaction with the mW particles.

\newpage

\section{Order Parameters used for Identifying Cluster Sizes}
\label{subsec:Identifying_Ice}

Clusters of ice-like particles were identified by first classifying
each particle in the system as either liquid-like or crystal like,
using the Steinhardt parameters defined in the main paper:
\begin{equation}
  \label{Eqn:SISteinhardt_barq6}
  \bar{Q}_6^{(i)} = \frac{Q_{6}^{(i)} + \sideset{}{^\prime}\sum_{j} Q_{6}^{(j)}}{\nu^{(i)}+1},  
\end{equation}
with $Q_6^{(i)}$ defined with the 6$^{\mathrm{th}}$ order spherical 
harmonics:
\begin{equation}
\label{Eqn:SISteinhardt_Q6}
  Q_6^{(i)} = \frac{1}{\nu^{(i)}}\sqrt{ \sum_{m=-6}^{6}
  \sideset{}{^\prime}\sum_{j,k} Y_{6m}^{\ast} (\hat{\mbf{r}}_{ij})Y_{6m} (\hat{\mbf{r}}_{ik})},
\end{equation}
For all mW simulations,
particle $j$ was considered adjacent (and so included in the summation
of Eqns.~\ref{Eqn:SISteinhardt_barq6} and~\ref{Eqn:SISteinhardt_Q6}) if
the distance between particles $i$ and $j$ $|\textbf{r}_{ij}| \leq
3.5\,$\AA. For this we used the implementation in Plumed\cite{Bonomi2019,
Tribello2014}, with the switching function $\sigma (|\textbf{r}_{ij}|)$:
\begin{equation}
    \label{Eqn:SISwitching_Function}
    \sigma (|\textbf{r}_{ij}|) = 
    \begin{cases}
        1 & \text{if }|\textbf{r}_{ij}| \leq 3.5\,\text{\AA} \\
        \exp \left( - \frac{(|\textbf{r}_{ij}| - 
        3.5\,\text{\AA})^{2}}{2\times (0.05\,\text{\AA})} \right) & 
        \text{if }3.5\,\text{\AA} \leq|\textbf{r}_{ij}| < 
        3.51\,\text{\AA} \\
        0 & \text{if }  3.51\,\text{\AA} \leq
        |\textbf{r}_{ij}|\\
    \end{cases}
\end{equation}
This switching function leads to the selection of the $\nu^{(i)}$
nearest atoms to particle $i$, where $\nu^{(i)}$ is defined as
\begin{equation}
    \nu^{(i)} = \sum_{j \neq i} \sigma 
    (|\textbf{r}_{ij}|)
\end{equation}
for all $j$ mW particles in the system.

\begin{figure}[h]
    \centering
    \includegraphics[width=\textwidth]{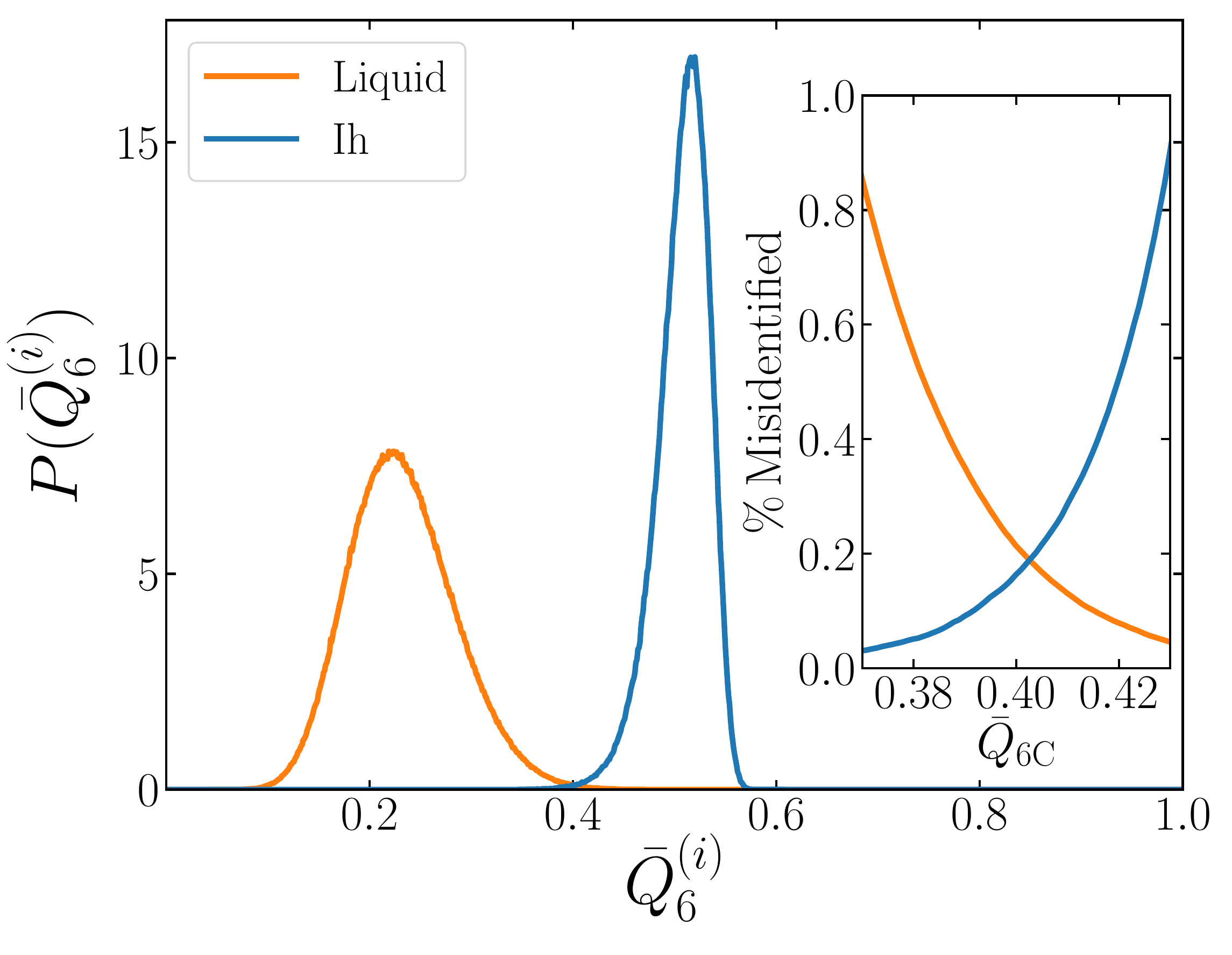}
    \caption{$P(\bar{Q}_{\rm 6}^{(i)})$ for both liquid and hexagonal 
    ice. Inset: The left hand side (blue) and right hand side (orange) 
    of Eqn.~\ref{Eqn:P(Q)} shown as a function of $P(\bar{Q}_{\rm 
    6C})$.} 
    \label{fig:q6_Cutoff}
\end{figure}

Once $\bar{Q}_{6}^{(i)}$ is identified for each mW particle $i$ in the
system, particles were identified as being ice-like or liquid-like
through the use of a cutoff value $\bar{Q}_{\rm 6C}=0.4021$.
$\bar{Q}_{\rm 6C}$ was found by obtaining the probability
distributions $P(\bar{Q}_{6}^{(i)})$ for both liquid mW and hexagonal
ice mW at 220\,K and 0\,bar,\footnote{Cubic ice was also considered,
and the corresponding distribution was found to be skewed to higher
values of $P(\bar{Q}_{6}^{(i)})$. Since all systems are being seeded
purely with hexagonal ice, we choose to base of derivation of
$\bar{Q}_{\rm 6C}$ on the liquid and Ih phases. } shown in
Fig.~\ref{fig:q6_Cutoff}.  $\bar{Q}_{\rm 6C}$ was found as the value
for which an equal proportion of Ih and liquid particles are
misidentified.\cite{Sanz2013,Espinosa2016}
\begin{equation}
    \int_{0}^{\bar{Q}_{\rm 6C}} P(\bar{Q}_{\rm 6,\,Hex}^{(i)}) d\bar{Q}_{\rm 6,\,Hex}^{(i)}=
    \int_{\bar{Q}_{\rm 6C}}^{1} P(\bar{Q}_{\rm 6,\,Liq}^{(i)}) d\bar{Q}_{\rm 6,\,Liq}^{(i)}
    \label{Eqn:P(Q)}
\end{equation}

Once all particles have been classified as liquid-like or
crystal-like, the largest cluster of adjacent crystal-like particles
is identified, using a depth-first clustering search and identifying
particles $i$ and $j$ as adjacent if $|\textbf{r}_{ij}| \leq
3.5\,\text{\AA}$.  The number of molecules of this cluster is taken as
$n_\mathrm{cl}$. Performing a similar analysis for TIP4P/ice at 
230\,K (with
$|\mbf{r}_{ij}|$ the distance between oxygen atoms on molecules $i$
and $j$ etc) gives $\bar{Q}_{\rm 6C} = 0.3946$.

\newpage

\section{Sensitivity of the nucleation rate from seeding to $\mathbf{\Delta \mu}$}

%% Here we present Fig.~3d. from the main text, recalculated with
%% $\Delta\mu$ at extreme values of the given literature range. We show
%% that our results are virtually indistinguishable, and our conclusions
%% remain consistent across the entire range.

\begin{figure}[H]
    \centering
    \includegraphics[width=0.5\textwidth]{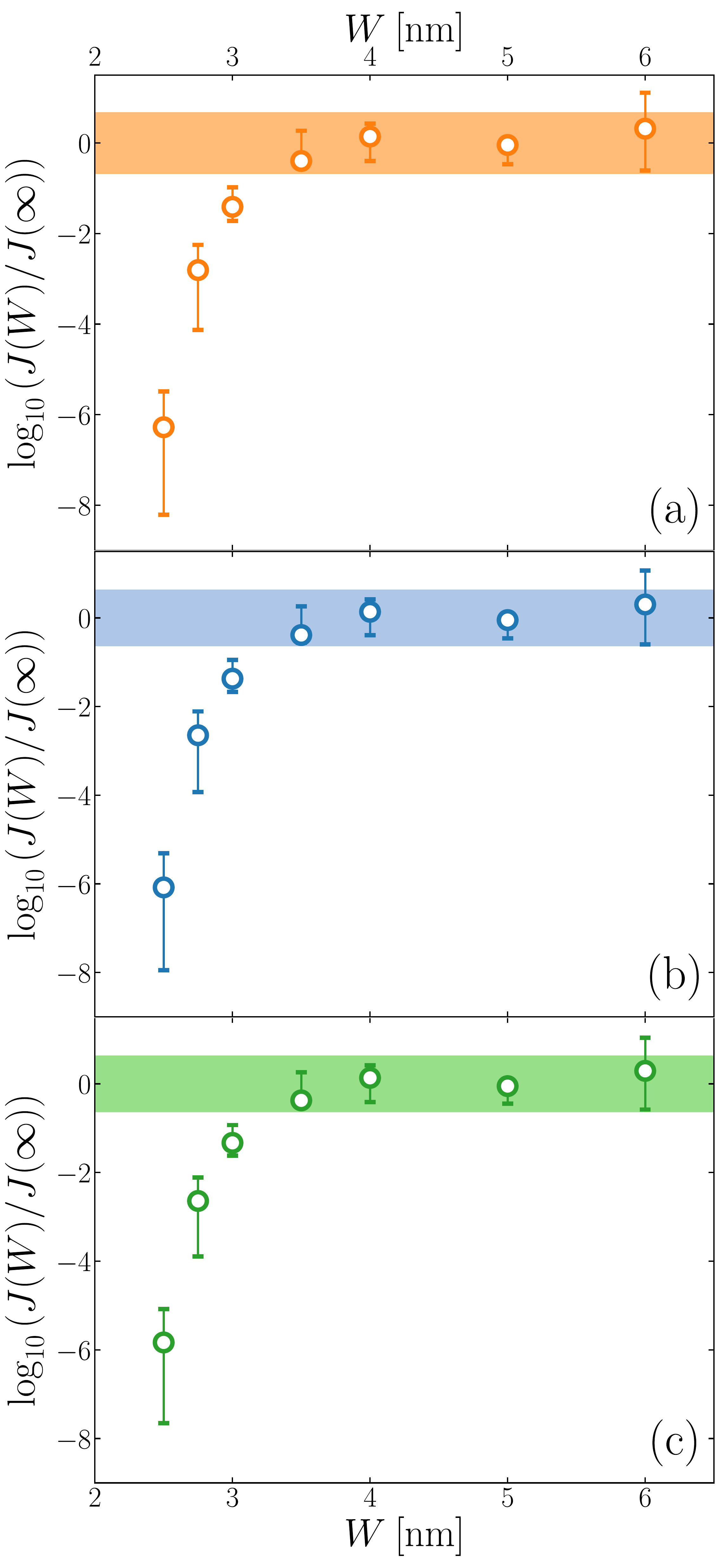}
    \caption{$J(W)$ obtained with (a) $|\Delta\mu|/k_{\rm B} =
      126$\,K, (b) $|\Delta\mu|/k_{\rm B} = 122$\,K (same as Fig.~3d
      in the main text), and (c) $|\Delta\mu|/k_{\rm B} = 118$\,K. The
      results are virtually indistinguishable from each other.}
    \label{fig:Delta_mu_Compared}
\end{figure}

\newpage

\section{Committor distributions of 2.5\,$\mathbf{nm}$ and 5\,$\mathbf{nm}$ films}

To test the suitability of $n_\mathrm{cl}$ as the ``reaction 
coordinate'', we present the committor distribution for 
$n_\mathrm{cl}$. The committor distribution was found by first 
selecting clusters for which $n_\mathrm{cl} (t=0) \approx 
n_\mathrm{cl}^\ddagger$ (taken in this case as $n_\mathrm{cl} (t=0) = 
n_\mathrm{cl}^\ddagger \pm 3$). This resulted in 72 initial clusters for 
$W=2.5\,\mathrm{nm}$ and 53 for $W=5\,\mathrm{nm}$. For each cluster, 
20 different simulations were performed where the velocities of each 
particle was randomized to create the expected velocity distribution at 
220\,K, and allowed to evolve until the seed grew or shrank. The 
probability $P(p_{\mathrm{B}})$ that a given cluster would grow to 
encompass the entire system is presented below.

\begin{figure}[h]
    \centering
    \includegraphics[width=\textwidth]{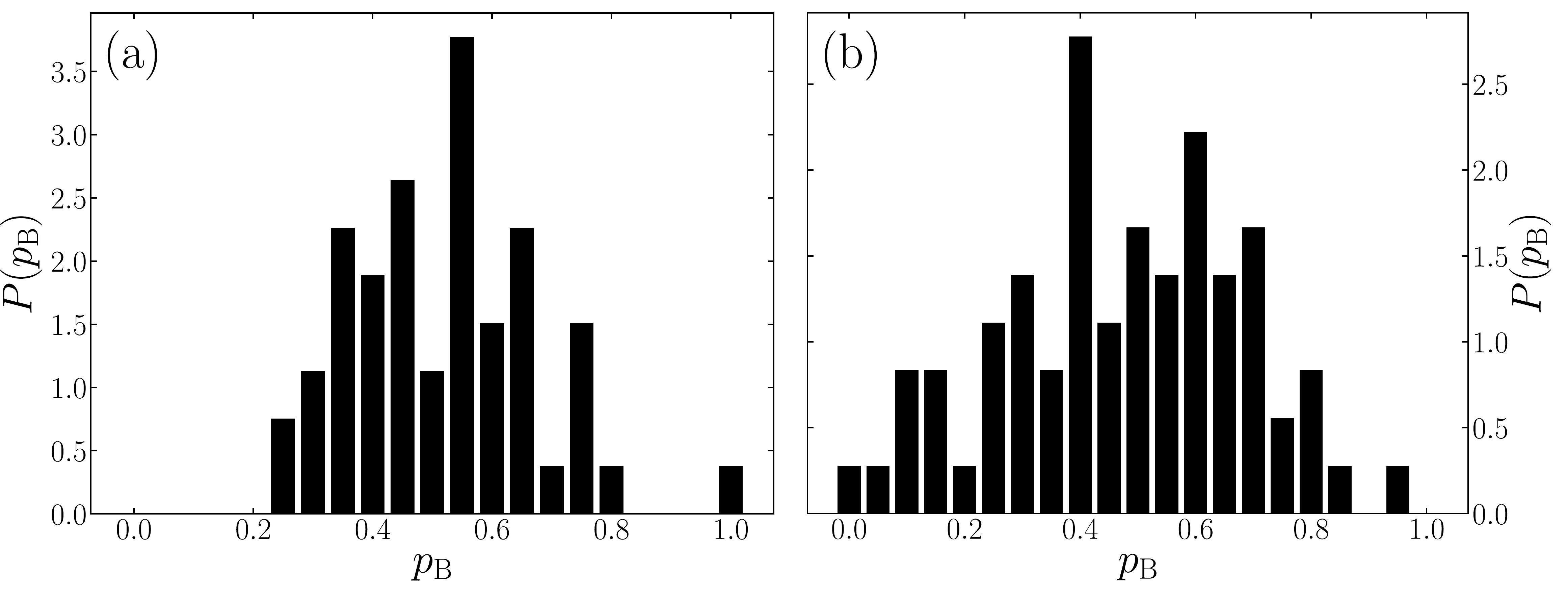}
    \caption{Committor distributions for (a) 5\,nm and (b) 2.5\,nm films.} 
    \label{fig:Committors}
\end{figure}

\newpage

\section{Density profiles for different values of $H$}

\begin{figure}[h]
    \centering
    \includegraphics[width=\textwidth]{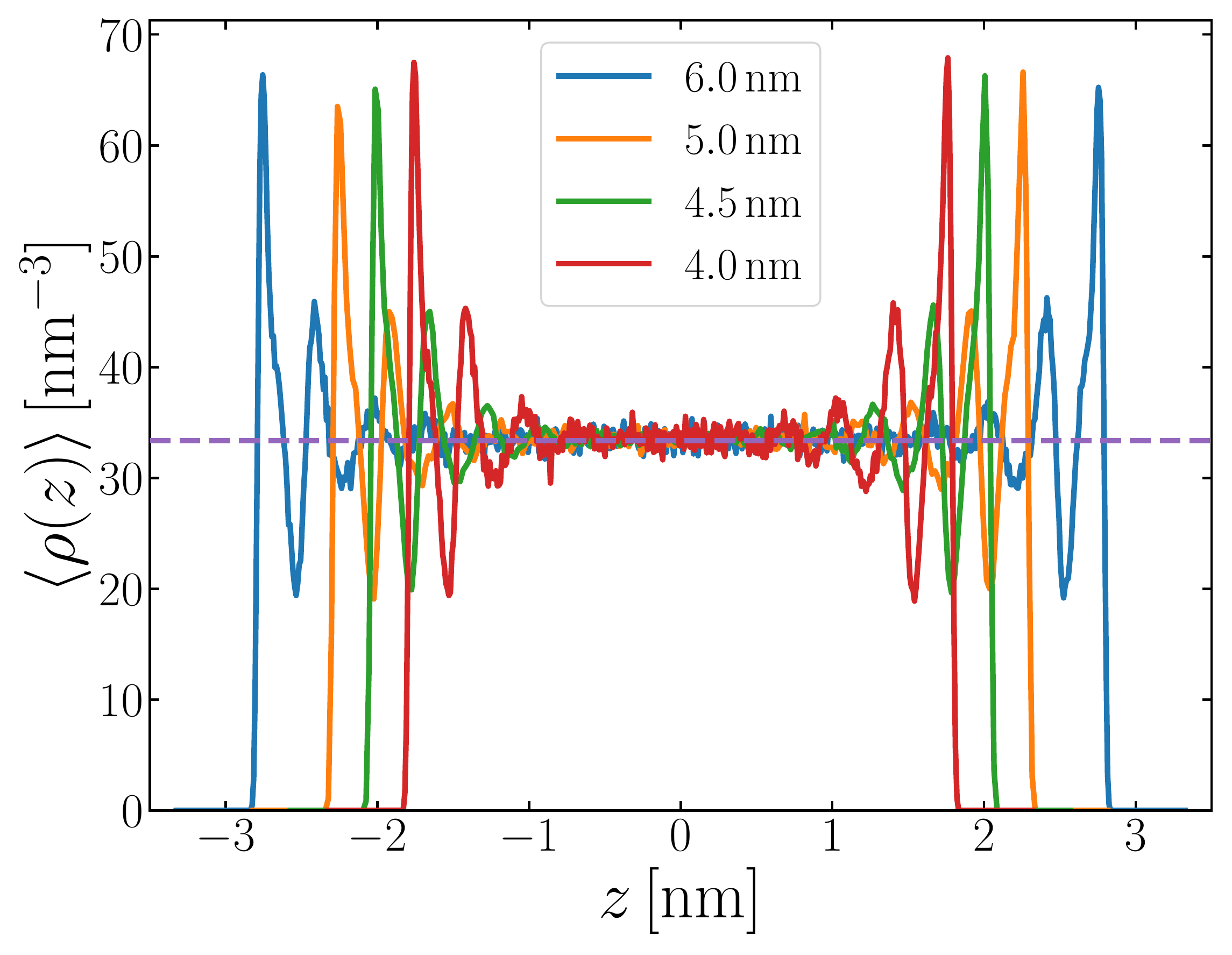}
    \caption{$\langle \rho(z)\rangle$ for $H=6.0\,\mathrm{nm},\,
      H=5.0\,\mathrm{nm},\,H=4.5\,\mathrm{nm},\,H=4.0\,\mathrm{nm}$.
      $\bar{\rho}=33.3774\,\mathrm{nm^{-3}}$, the bulk density for mW
      at 220\,K and 0\,bar, is included as the dashed line.}
    \label{fig:SI_Density_Profiles}
\end{figure}

\newpage

\section{Volumetric nucleation rates obtained from cooling ramps}

Volumetric nucleation rates for the cooling ramps investigation, found
by dividing the freezing rates of the bulk system by the $Nv$ where
$N=6000$ is the number of molecules in the system, and $v=3 \times
10^{-29}\,\mathrm{m^{-3}}$ is the volume of a single mW molecule under
these conditions.  These results are compared to previous literature
values.\cite{HajiAkbari2018,russo2014new,moore2011structural,
  Sanchez2022}

\begin{figure}[h]
    \centering
    \includegraphics[width=\textwidth]{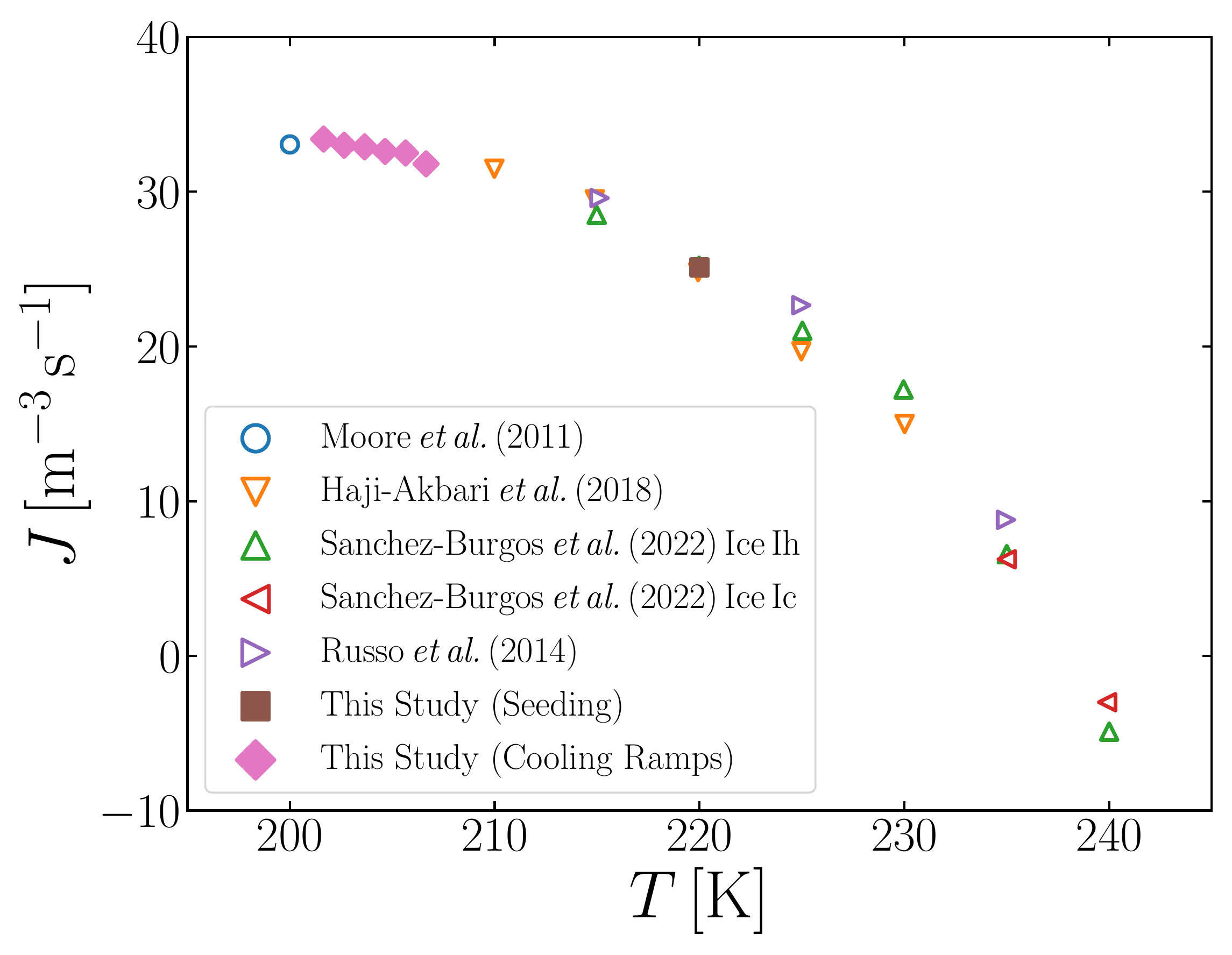}
    \caption{$J(\infty)$ for this work and a range of previous 
    studies.} 
    \label{fig:Cooling_Rates}
\end{figure}

\newpage

\section{Preliminary Work on TIP4P/ice Films}
\label{subsec:TIP4Pice_Seeding}

We argue in the main paper that mW is suitable for this study, 
due to the mechanisms of ice formation being similar to TIP4P/ice$^{(8.5)}$, 
and similar convergence of structural order to bulk-like values within $\sim 1\,$nm from 
the surface. In this section, we present preliminary work using the 
TIP4P/ice$^{(8.5)}$ model for both bulk and film ($W=4.3\,\mathrm{nm}$).
Seeding was implemented in a manner similar to 
Sec.~\ref{subsec:mW_Seeding}, but at 230\,K and a 200\,ps equilibration time 
instead of 80\,ps. Additionally, the same method was used to identify 
ice-like molecules, with the oxygen in each molecule acting as the 
point which $\textbf{r}_{ij}$ vectors point to and from. 
Due to the different conditions and model, $\bar{Q}_{\rm 6C}=0.3946$.

\begin{figure}[h]
    \centering
    \includegraphics[width=\textwidth]{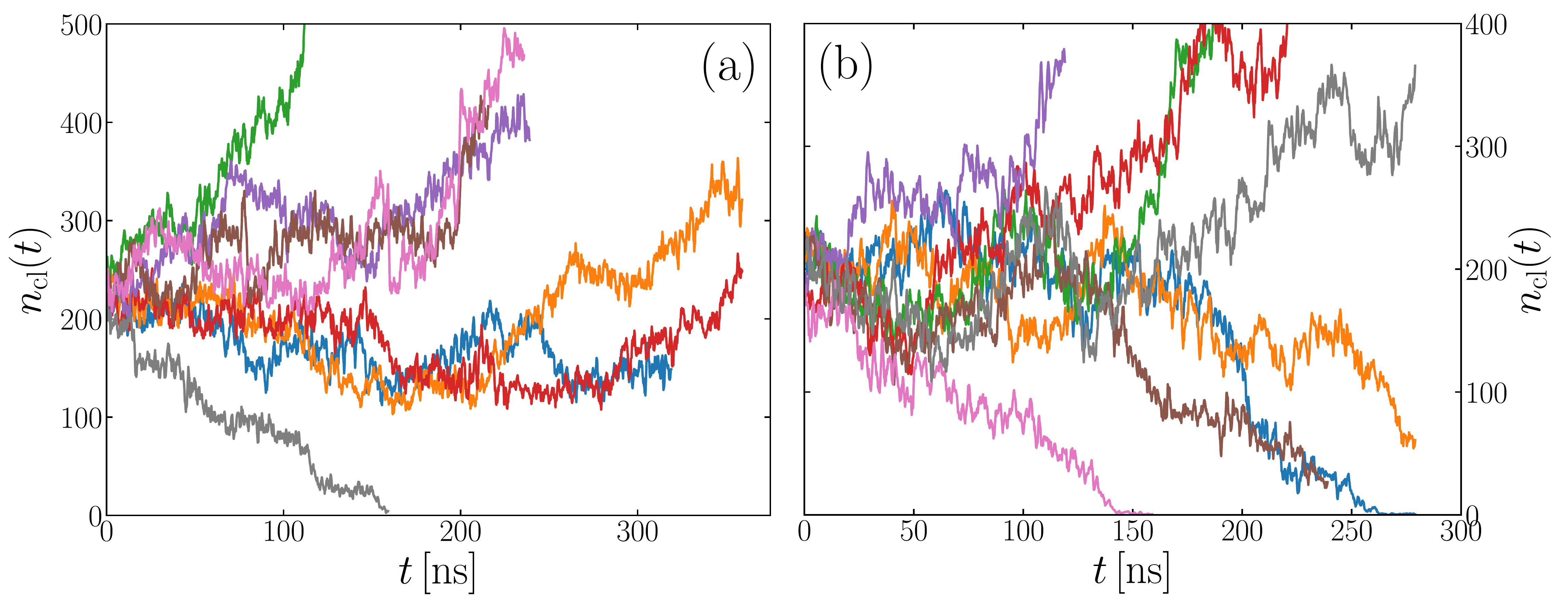}
    \caption{$n_{\mathrm{cl}}(t)$ for 8 independent seeding simulations in 
    a) a bulk system, $\langle n_{\mathrm{cl}}(t=0) \rangle = 232$ and b) a 4\,nm 
    film $\langle n_{\mathrm{cl}}(t=0) \rangle = 200$. In both cases, clusters of
    similar initial size grow and shrink, and some clusters spend significant 
    time at or around their initial sizes. Both facts indicate these systems are 
    at or near criticality.} 
    \label{fig:TIP4PICE}
\end{figure}

This investigation was extremely limited in scope due to the high computational 
cost of TIP4P/ice$^{(8.5)}$. However, we show that both film and bulk appear 
to be near critical for $n_\mathrm{cl}(t=0) \approx 200$. The fact that 
bulk and film systems appear to report similar critical nuclei further supports 
our mW-based study.

\end{document}